\documentclass[11pt]{article}
\usepackage{amsmath}
\usepackage{amsthm}
\usepackage{epsfig}
\usepackage{boxedminipage}
\usepackage{auto-pst-pdf}
\usepackage{graphicx}
\usepackage{pstricks}

\newcommand{\ds}{d^\star}
\newcommand{\bone}[3]{{\bf 1}^{#1}_{#2,#3}}
\newcommand{\proc}[3]{{P^{#1}_{#2,#3}}}

\newcommand{\rw}{w^r}
\newcommand{\wf}{w^f}
\newcommand{\dr}{d^r}

\newcommand{\wmin}{w_{\min}}
\newcommand{\dsr}{{d^{\star}}}

\newcommand{\eat}[1]{}
\newcommand{\eps}{{\varepsilon}}

\newcommand{\jrej}{{J^{R}}}
\newcommand{\opt}{{T^\star}}
\newcommand{\load}{{\tt load}}
\newcommand{\maxload}{{\tt maxload}}
\newcommand{\level}{{\tt level}}

\newtheorem{theorem}{Theorem}[section]

\newtheorem{lemma}[theorem]{Lemma}

\newtheorem{claim}[theorem]{Claim}

\newtheorem{corollary}[theorem]{Corollary}

\newcommand{\intunit}[2]{{{\cal I}^{({#1},{#2})}}}
\newcommand{\intg}[3]{{{\cal I}^{({#1}, {#2}, {#3})}}}

\newcommand{\ints}[1]{{{\cal I}^{(#1)}}}
\renewcommand{\sp}{\hspace*{2 mm}}
\newcommand{\length}{{\tt length}}
\newcommand{\len}[1]{{L^{(#1)}}}
\newcommand{\jobs}[1]{{J^{(#1)}}}
\newcommand{\jobg}[2]{{J^{(#1,#2)}}}

\newcommand{\A}{{\cal A}}
\newcommand{\B}{{\cal B}}
\newcommand{\I}{{\cal I}}

\newcommand{\logeps}{{\log \left( \frac{1}{\eps} \right)}}
\newcommand{\logsqeps}{{\log^2 \left( \frac{1}{\eps} \right)}}

\newcommand{\bI}{{\bf I}}

\newcommand{\powd}[1]{{2^{#1-\ds}}}
\newcommand{\loadbal}{{\tt LoadBalancing}}
\newcommand{\linf}{{\tt MaxFlowTime}}
\newcommand{\wtdlinf}{{\tt WtdMaxFlowTime}}
\renewcommand{\O}{{\cal O}}
\newcommand{\bint}[3]{{\bI}^{(#1,#2,#3)}}
\newcommand{\type}{{\tt type}}
\newcommand{\is}{{i^\star}}
\newcommand{\jobB}[2]{{J^{\A}_{#1,#2}}}
\newcommand{\jobrejB}[2]{{J^{\tt rej}_{#1,#2}}}

\newcommand{\wtdlinfgen}{{\tt GenWtdMaxFlowTime}}

\newcommand{\footremember}[2]{%
   \footnote{#2}
    \newcounter{#1}
    \setcounter{#1}{\value{footnote}}%
}
\newcommand{\footrecall}[1]{%
    \footnotemark[\value{#1}]%
}

\begin{document}

\title{Rejecting jobs to Minimize Load and Maximum Flow-time}
\date{}

\author{%
    Anamitra Roy Choudhury\footremember{ibm}{IBM Research - India, New Delhi. \ email: {\tt anamchou@in.ibm.com}}%
    \and Syamantak Das\footremember{iitd}{Department of Computer Science and Engg., IIT Delhi. \  email: {\tt \{sdas, naveen, amitk\}@cse.iitd.ac.in }}%
    \and Naveen Garg\footrecall{iitd} %
    \and Amit Kumar\footrecall{iitd} %
}




\maketitle
\begin{abstract}
Online algorithms are usually analyzed using the notion of competitive ratio which compares
the solution obtained by the algorithm to that obtained by an online adversary for the worst
possible input sequence. Often this measure turns out to be too pessimistic, and one popular
approach especially for scheduling problems has been that of  ``resource augmentation" which was
first proposed by Kalyanasundaram and Pruhs. Although resource augmentation has been very successful in dealing
with a variety of objective functions, there are problems for which even a (arbitrary) constant
speedup cannot lead to a constant competitive algorithm. 
In this paper we propose a  ``rejection model" which requires no resource
augmentation but which permits the online algorithm to not serve an epsilon-fraction of the requests.

The problems considered in this paper are in the restricted assignment setting where each job
can be assigned only to a subset of machines. For the load balancing problem where the objective
is to minimize the maximum load on any machine, we give $O(\log^2 1/\eps)$-competitive algorithm
which rejects at most an $\eps$-fraction of the jobs. For the problem of minimizing the maximum
weighted flow-time, we give an $O(1/\eps^4)$-competitive algorithm which can reject at most an $\eps$-fraction of the jobs by weight. 
We also extend this result to a more general setting
where the weights of a job for measuring its weighted
flow-time and its contribution towards
total allowed rejection weight are different. This is useful, for instance, when we consider the objective of minimizing the maximum stretch.
We obtain an $O(1/\eps^6)$-competitive algorithm in this
case.

Our algorithms are immediate dispatch, though they may not be immediate reject. All these
problems have very strong lower bounds in the speed augmentation model.

\end{abstract}
\thispagestyle{empty}
\pagebreak
\setcounter{page}{1}


\section{Introduction}
Online algorithms are usually analyzed using the notion of competitive ratio which compares the solution obtained by the algorithm to that obtained by an offline adversary for the worst possible input sequence. Researchers have tried to address the criticism of this measure being too pessimistic by either limiting the power of the adversary -- oblivious adversary, stochastic adversary -- or giving more power to the online algorithm -- lookahead, additional resources, etc. One popular approach especially for scheduling problems has been that of ``resource augmentation'' and was first proposed by  Kalyanasundaram and Pruhs~\cite{KalyanasundaramP95}. In this model the machines of the online algorithm have more speed than those of the offline algorithm. Many scheduling problems for which no constant competitive online algorithm is possible now have such algorithms in this resource augmentation model.  The success of the speed augmentation model lies in the fact that many natural algorithms can be analysed in 
this framework.


In this paper, we propose a ``rejection model'' in which there is no resource augmentation, but we allow the online algorithm to 
not serve an $\eps$-fraction of requests. There are two principal reasons for considering this model: (i) although resource augmentation has been very successful in dealing with a variety of objective functions, there are problems for which even a (arbitrary) constant speedup cannot lead to a constant competitive algorithm -- we consider two such problems in this paper, and (ii) this might be
a natural assumption in many settings where job rejection is part of the service provided by a system (e.g., 
``Server busy: Please try again later'' message we often see when accessing popular websites).



For most scheduling problems, an algorithm in the resource augmentation model can be ``simulated" in this rejection model by roughly letting each machine drop every $(1/\eps)^{th}$ job assigned to it. However, the rejection model is much more powerful than the resource augmentation model since we are not restricted to drop an $\eps$-fraction of jobs assigned to each machine, i.e., we could drop many more jobs assigned to one machine as compared to another as long as the overall number
of rejected jobs stays within $\eps$-fraction of the total number of jobs. We demonstrate this by considering two classical problems -- load balancing and maximum (weighted) flow time. For both these problems we give constant competitive algorithms in the rejection model while no such algorithms are possible in the resource augmentation model. This is the key contribution of this paper.

The problems considered in this paper are in the restricted assignment setting where each job can be assigned only to a subset of machines. All our algorithms are pre-emptive, immediate dispatch -- a job is assigned to a machine as soon as it is released and non-migratory -- a job is processed only on the machine to which it is assigned. While ideally we would also like to make rejection decisions immediately when the job is released, we also allow for the job being rejected while it is waiting in the queue. We justify this by showing that no online algorithm with immediate dispatch and immediate rejection can be constant competitive for the load balancing problem (and hence,
the maximum flow-time problem).

\noindent
{\bf Our Results}: For the load balancing problem(\loadbal) where the objective is to minimize the maximum load on any machine, we give an $O(\logeps)$ competitive immediate rejection algorithm when all jobs have unit processing time and the online algorithm can reject an $\eps$-fraction of the jobs. For general processing times our algorithm is not immediate reject and is $O(\logsqeps)$-competitive. We show that one cannot get a constant competitive algorithm if we require immediate rejection.  Note that there is a $\Omega(\log m)$ lower bound on the competitive ratio of any online algorithm for this problem where $m$ is the number of machines~\cite{Azar95}. Further, making the machines an $\eps$-fraction faster has no significant impact on this lower bound.

For the maximum flow-time problem Anand~et.~al~\cite{AnandBFGK13} show that no immediate dispatch algorithm can be constant competitive even when the jobs are unit length and we allow resource augmentation. For this setting of unit jobs we show an immediate reject algorithm which is $O(1/\eps)$-competitive and rejects at most an $\eps$ fraction of the jobs. When jobs have weights(\wtdlinf), the objective is to minimize the maximum weighted flow-time of a job. For this setting Anand~et.~al~\cite{AnandBFGK13} show that no online algorithm can be constant competitive even when we allow non-immediate dispatch and resource augmentation. Our algorithm for this setting is immediate dispatch but not immediate reject and is allowed to reject jobs of total weight at most an $\eps$-times the total weight of all jobs and has a competitive ratio of $O(1/\eps^4)$. We also show that it is not possible to get better than $O(1/\eps)$-competitive algorithm in this model, and  that one cannot get a good competitive algorithm 
if we are required to perform immediate reject and immediate dispatch. 

We further generalize our result to a setting where each job has two kinds of weight -- rejection weight and flow-time weight. The weighted flow-time of the job
is defined as its flow-time times its flow-time weight, and the goal, as before, is to minimize the maximum weighted flow-time of a job. However, the total 
rejection weight of the jobs which get rejected should be at most $\eps$ fraction of the total rejection weight of all the jobs. We obtain an $O(1/\eps^6)$-competitive algorithm for this problem. The problem of minimizing maximum stretch is a special case of this setting in the rejection model. Here, the rejection weights are all unit while the flow-time weights are the inverse of processing sizes.

\vspace*{-0.1in}
\section{Related Work}
{\noindent}{\bf Load Balancing}. Graham~\cite{Graham66} considered this problem in the context of identical machines and showed that the simple greedy
heuristic of assigning the next task to the least loaded machine is $2$-competitive (see also the survey by~Azar~\cite{AzarSurvey92}).  
Albers~\cite{Albers99} improved the competitive ratio to $1.923$ and also showed a lower bound of $1.852$ on the competitive ratio of  any deterministic online algorithm, while Albers~et~al.~\cite{Albers12} shows improved bounds in the special case where the online algorithm knows the sum of job sizes at any point.   
For related machines model, Berman~et~al.~\cite{Berman97} gave constant competitive algorithms. However, the problem becomes significantly harder 
in the unrelated machines model, where a job can have different processing time on different machines.
 Azar~et~al.~\cite{Azar95} considered the problem in the restricted assignment setting, and gave 
 an $O(\log m)$-competitive algorithm for load balancing($m$ being the number of machines). They also complemented this result 
 by proving lower bound of $\Omega(\log m)$ for any deterministic and $\Omega(\ln m)$ for any randomized online algorithm
  under the restricted assignment model. Buchbinder~et~al.\cite{Buchbind06} gave an 
alternative, more general upper bound on the load on any prefix of the most loaded machines. For the unrelated machines setting, Aspnes~et~al. \cite{Aspnes97} gave an $O(\log m)$-competitive algorithm. There has been some work on resource augmentation in this setting. 
Azar~et~al.\cite{Azar00} showed a competitive ratio of $1 + 1/2^{\frac{n}{m}(1-o(1))}$ when the online algorithm is allowed to use  $n$ identical machines while the offline optimal is restricted to $m < n$ identical machines. \\ 
{\noindent}{\bf Flow-time minimization}.  There has been considerable work on scheduling with the objective 
of minimizing a suitable norm of the flow-time of jobs. For the objective of average flow-time of jobs, 
 a logarithmic competitive algorithm in the identical machines setting is known~\cite{LeonardiR07, AvrahamiA03}. 
Garg and Kumar~\cite{GargK06} extended this result to the related machines setting. 
Garg and Kumar~\cite{GargK07} showed that the problem becomes considerably harder in the restricted assignment setting and no
online algorithm with
bounded competitive ratio is possible. Bansal and Pruhs~\cite{BansalP03} showed
that the competitive ratio can be as high as $\Omega(n^c)$  for the problem of minimizing  $\ell_p$ (for any $1<p<\infty$) norm,
where $n$ is the number of jobs, even for a single machine.
For minimizing the maximum flow-time in the identical machines model, Amb\"uhl and Mastrolilli \cite{AmbuhlM05} gave a simple 2-competitive algorithm. 
However, Anand et al.~\cite{AnandBFGK13}  showed that the competitive ratio of any online algorithm for the restricted assignment setting is 
as high as  $\Omega(m)$, where $m$ is the number of machines.


One approach for circumventing such strong lower bounds has been speed augmentation, where
 we allow each machine
in the online algorithm $\eps$-fraction more speed than the
corresponding machine in the offline algorithm. This model was first proposed by 
Kalyanasundaram and Pruhs~\cite{KalyanasundaramP95} who used it to get an $O(1/\eps)$-competitive algorithm
for minimizing total flow time on a single machine in the non clairvoyant setting.
Bansal and Pruhs~\cite{BansalP03}
 proved that several natural scheduling algorithms 
are $O(1/\eps)$-competitive algorithm for minimizing 
$\ell_p$ norm (for any $1 < p < \infty$) of flow-time of jobs in the  single machine setting. Golovin et~al.~\cite{GolovinGKT08}
extended this result to parallel machines setting. Chekuri~et~al. \cite{ChekuriGKK04} showed that the 
immediate dispatch algorithm of Avrahami and Azar~\cite{AvrahamiA03}
is also $O(1/\eps)$-competitive for all $\ell_p$ norms ($p \geq 1$). 

In the more general setting of unrelated machines with speed augmentation, 
Chadha~et~al. \cite{ChadhaGKM09} gave an $O(1/\eps^2)$-competitive algorithm for 
minimizing the sum of flow-time of jobs, which was improved and extended to the case of 
$\ell_p$ norm of flow-time by Im and Moseley~\cite{ImM11} and Anand~et~al.~\cite{AnandGK12}.
However, the competitive ratio remained a function of $p$, and in fact, Anand et al.~\cite{AnandBFGK13} showed that
one cannot obtain competitive ratio better than $\Omega \left( \frac{p}{\eps^{1-O(1/p)}} \right)$ even in the restricted assignment model. 

For the objective of minimizing the maximum (unweighted) flow time on unrelated machines under the  speed augmentation model,
Anand~et~al.~\cite{AnandBFGK13} gave a $O(1/\eps)$-competitive algorithm; however their algorithm is not an immediate dispatch 
algorithm. In fact, Azar et al. \cite{Azar95} showed that any immediate dispatch
algorithm for minimizing maximum flow time
will have a competitive ratio can be as high as $\Omega(\log m)$ in the restricted assignment setting even with constant
speed augmentation. In the maximum weighted flow-time case, this lower bound holds even if we  allow non-immediate
 dispatch~\cite{AnandBFGK13}. Chekuri et al.~\cite{ChekuriM09, ChekuriIM12} considered the problem of minimizing the maximum delay factor of a job in
the parallel machines setting,
where jobs come with deadlines and the delay factor of a job is the ratio of its flow-time to the difference between 
the deadline and its release date. 
This problem is 
equivalent to minimizing the maximum weighted flow time, where the weight of a job $j$ can be seen as
$(d_j-r_j)^{-1}$.  They gave a $(1+\eps)$-speed $O(1/\eps)$-competitive algorithm.  \\
{\noindent}{\bf Scheduling with Rejection.} There has been considerable work on online scheduling with job rejections in the prize collecting setting. Here, each job comes with a specified penalty  which is to be incurred in case it is not scheduled. The goal is to minimize the sum of a suitable objective function of the completion times of the jobs (which are not rejected) and the total penalty cost of rejected jobs. Bartal~et~al.~\cite{BartalLMSS00} considered the problem of minimizing makespan in the identical machines model with penalties and gave a 2.618-competitive algorithm. Epstein et al.~\cite{Epstein14} extended this work to the problem of minimizing makespan on two related machines. 
Bansal et al.~\cite{Bansal03} considered the online problem of minimizing total flow-time of jobs and total idle time on a single machine along with uniform rejection penalty for all jobs. They gave 2-competitive algorithms for both the objectives. For the case of arbitrary penalties and average weighted flow-time, they showed strong lower bound of $\Omega(\max(n^{\frac{1}{4}}, C^\frac{1}{2}))$ on the
competitive ratio of  any randomized online algorithm. They complemented this with an $O(\frac{1}{\eps}(\log W + \log C)^2)$-competitive  algorithm
with $(1+\eps)$-speed augmentation, where $W$ is the ratio between maximum and minimum weights and $C$ is the ratio between maximum and minimum penalties. These results were extended to non-clairvoyant settings (i.e., the size of a job is not known till the time it finishes processing) in the identical machines setting
by Chan~et~al.~\cite{Chan12} -- they considered an objective function which also had an energy term, and gave 
 constant competitive $2(1+\eps)$-speed algorithm. Minimizing average flow-time and completion time with rejections has also been studied in the offline context~\cite{Gupta09}.

Charikar and Khuller~\cite{Charikar06} studied the online problem of minimizing maximum flow-time in  more general context of broadcast scheduling, where both the online algorithm and the offline optimum are allowed to ignore a fixed fraction of the jobs. They showed that no randomized online algorithm can be constant competitve, while the offline problem admits a 5-approximation.

\section{Problem Statement}
\label{secf:statement}
We formally define the problems considered in this paper. We consider the online problem of scheduling in the restricted assignment setting. The input instance specifies a small enough positive parameter $\eps$.  We have a set of $m$ machines, and jobs arrive in an online manner.
 A job $j$ arrives at time $r_j$, and can only be scheduled on a subset $S_j$ of machines. Further, it specifies a processing requirement (or size) of $p_j$ units.

In the {\loadbal} problem, a solution needs to dispatch each job to a machine. It can also choose to reject a job (either when it arrives or after a job has been dispatched to a machine). However, for any time $t$, the total number of jobs rejected by the algorithm till time $t$ must be at most $\eps$ fraction of the number of jobs that have arrived till time $t$.
The {\em load} assigned to a machine at any point of time $t$  is defined as the total processing requirement of jobs which have been dispatched to it till time $t$ -- note that this does not count jobs which get rejected by time $t$, but it counts jobs which will get rejected after time $t$.
The goal of the scheduling algorithm is to minimize (for all time $t$) the maximum load assigned to any machine.

In the {\wtdlinf} problem, an input instance can be described as above. Further, we also have a weight $w_j$ associated with each job $j$. 
A solution needs to process a job $j$ on one of the machines in $S_j$ for $p_j$ amount of time. Note that any machine can perform 1 unit of processing in unit time, and we allow jobs to be pre-empted. However, migration is not allowed -- a job dispatched to a machine $i$ must be completed on $i$ or rejected. 
The flow-time of a job is defined as the difference between its completion time $C_j$ and its release time $r_j$. The goal is to minimize the maximum weighted
 flow-time of a job, i.e., $\max_j w_j \cdot (C_j - r_j)$.
 As before, till any time $t$,  our algorithm is allowed to reject $\eps$-fraction of jobs which have arrived till this time. This rejection can either happen on arrival of a job or later.

In the {\wtdlinfgen} problem, the setting is same as above, but 
 the weights for flow-time and the weight for rejection are different. In other words, instead of having one weight
$w_j$, a job $j$ has two weights associated with it, the rejection-weight $w^r_j$ and flow-time-weight $w^f_j$. Again the objective is to minimize the maximum 
over all jobs $j$ of $w^f_j F_j$, where $F_j$ denotes the flow-time of job $j$ in a schedule; and we are allowed to reject jobs of total rejection-weight at most
$\eps$ times the total rejection-weight of all the jobs. Note that in the case of {\wtdlinf} problem, $w^r_j$ happens to be same as $w^f_j$.

Our algorithms for {\linf} and {\wtdlinf} (and {\wtdlinfgen})
 satisfy the  immediate dispatch property -- when a job $j$ arrives, it is dispatched  to a machine at time $r_j$. Note that it may still get rejected later.

We now give an outline of rest of the paper.  
In Section~\ref{secf:techniques}, we give a brief outline of our algorithms and the ideas involved. 
In Section~\ref{secf:loadbal},
we give an algorithm for the {\loadbal} problem.  In Section~\ref{secf:linf}, we describe our algorithm for the {\wtdlinf} problem. To illustrate some of the ideas, 
we first consider the special case of unit size and unit weight in Section~\ref{secf:unitsize}. Although one does not need to invoke an LP relaxation here, we 
give a more detailed explanation in this section so that similar ideas in later section can be clearer. In Section~\ref{secf:gensize}, we describe the algorithm 
for the general case. The algorithm is split in two parts -- $\A$ and $\B$. We first describe these two algorithms and give their analysis in Section~\ref{secf:analysis}. 
In Section~\ref{secf:extension}, we show how to extend our results to the {\wtdlinfgen} problem. Finally, 
in  Section~\ref{secf:lower}, we give lower bounds on competitive ratios for the problems considered in this paper.

\section{Our Techniques}
\label{secf:techniques}
In both the {\loadbal} and the {\wtdlinf} problems, we assume that we know the value of the offline optimum solution, denoted by $\opt$. Removing
this assumption requires standard ideas in online algorithms (details are given in Section~\ref{secf:remload} and Section~\ref{secf:remlinf}).

In the {\loadbal} problem, we first consider the special case when all jobs are of unit size. Here, the algorithm is quite natural -- when a job
arrives, it is dispatched to the least loaded machine (among the ones it can be dispatched to). However, we do not allow the load
on a machine to exceed $\alpha \opt$ where $\alpha=\logeps$ -- if we dispatch a job to a machine on which the load is already at this limit, we reject this job.
Clearly, the algorithm is $O(\logeps)$-competitive. To bound the number of rejected jobs, we use the following argument. Let $m_l$ denote the
number of machines where the algorithm dispatches at least $l \opt$ jobs ($l$ is some integer between 1 and $\logeps$). A naive argument will show 
that the total number of jobs  is at least $m_l \cdot l\opt$, but a slightly more careful argument can  show that the number of jobs in the system is at least $m_l \cdot 2^l \opt$. A rough argument goes as follows: let $M_l$ be the set of machines where the algorithm
assigns at least $l \opt$ jobs. 
 Consider the jobs dispatched to $M_l$ which are at level $(l-2) \opt$ or above (level of a job is its position in the queue of the corresponding machine). There will be $2 |M_l| \opt$ such jobs, and so there must be a set of $2 |M_l|$ machines where such jobs can be processed (otherwise
optimum value cannot be $\opt$). Each of these machines must have  at least $(l-2) \opt$ jobs assigned to them (because of the job dispatch policy). 
Thus, the total number
of jobs is at least $2(l-2)|M_l| \opt$. Continuing this argument gives the exponential scaling with $l$. Now, if the number of 
rejected jobs is large, then there must be many machines where the load is up to the maximum limit, which means that there must be a huge
(exponential in $\alpha$ times number of such machines) number of jobs in the system.

For the more general case of the {\loadbal} problem when job sizes are arbitrary, we can reduce the problem to the case of unit size jobs.
Assume wlog that all job sizes are powers of 2. For each value of index $j$, we run the above algorithm independently for jobs of size $2^j$.
This will ensure that for any fixed $j$, the total load (i.e., processing time) of jobs of size $2^j$  assigned to a machine is at most
$\alpha\opt$. In case this limit is reached on a machine for many values of the index $j$, one can show that we can reject many (large) jobs.
Note that this algorithm may reject a job after it gets dispatched to a machine. We show that this is unavoidable (details in Theorem~\ref{thmf:lowerunbounded}).

In the {\wtdlinf} problem, each job $j$ has a weight $w_j$ and the goal is to minimize the maximum over all jobs $j$, of $w_j F_j$, where 
$F_j$ is the flow-time of $j$. In this problem, 
the ideas are more subtle and technically involved. In fact, we show that unlike the {\loadbal} problem, one
cannot obtain better than $O(1/\eps)$-competitive algorithm (details in Theorem~\ref{lemf:lower-linf}). 
To give intuition about our algorithm and analysis techniques, we consider some
special cases:

\noindent
{\bf Unit size and unit weight:} Suppose all jobs are of size 1 and weight 1. In this case a natural algorithm is as follows: each machine
maintains a queue of jobs assigned to it. When a new job arrives, it is dispatched to the machine with the smallest queue (among the machines
to which it can be processed on). However, if all such machines have at least $\opt/\eps$ jobs in their queue, we reject this job. The analysis
is somewhat trickier than the corresponding case for \loadbal. The reason is as follows: suppose there is a job $j$ which can go on two machines $i_1$ and $i_2$, and both have the same queue size when $j$ is released. Suppose we dispatch $j$ to $i_1$. But in future, $i_1$ will continue
to get more jobs assigned to it (because perhaps these jobs could only get processed on $i_1$), whereas $i_2$'s queue will decrease with time.
In the {\loadbal} problem, queues do not dissipate with time, and so the proof gets much simpler.


\begin{figure}
\begin{center}
\input{interval.pstex_t}
\caption{Example showing intervals $\intunit{i}{l}$ for a machine $i$. Here $\intunit{i}{1}=\{I^1_1\}, \intunit{i}{2}=\{I^1_2\}, \intunit{i}{3}=\{I^1_3,
I^2_3, I^3_3 \}, \intunit{i}{4}=\{I^1_4\}, \intunit{i}{5}=\{I_5^1\}$}
\label{figf:interval}
\end{center}
\end{figure}

Our proof idea is as follows. For each machine $i$ and parameter $l$ ($l$ varies between 1 and $1/\eps$), we define a set of disjoint intervals $\intunit{i}{l}$. 
An interval $I$ in this set is a minimal interval such that the queue size in $i$ at the left end-point is $(l-1) \opt$ and that at the right end-point is 
$l \opt$ (see Figure~\ref{figf:interval}).  For different values of $l$ and fixed $i$, these intervals form a laminar family. Let $m_l$ be the number of intervals in 
$\cup_i \intunit{i}{l}$. Suppose the algorithm rejects $k \opt$ jobs. Then it is not difficult to show that 
\begin{eqnarray}
\label{eqf:example}
m_{l-1} \opt + L_{l-1}\geq m_{l} \opt + L_l + k \opt,
\end{eqnarray}
 where $L_l$ denotes the total length of all the intervals in $\cup_i \intunit{i}{l}$. Indeed, 
the RHS above denotes a set of jobs which get released during an interval in $\cup_i \intunit{i}{l}$, and each such interval can only take $\opt$ jobs beyond its 
own length. Summing the above for all $l$ between 1 and $1/\eps$ shows that the number of jobs is at least $k \opt/\eps$, and so the algorithm rejects
only $\eps$ fraction of the jobs. The formal proof is given in Section~\ref{secf:unitsize}. Although one can simply formalize the argument given above, we 
give a somewhat longer proof which sets dual variables for a natural LP relaxation of this problem.  This proof generalizes to the more general case.

In more general settings, we split the algorithm in two parts: the first part ensures that the queues on each machine are bounded, and the second part
uses this property to ensure that all jobs which do not get rejected finish within the required time. We describe details of the first part below.

\noindent
{\bf Unit weight and arbitrary size}  
For sake of simplicity, assume that job sizes are 1 or 2. Each machine $i$ maintains two queues: $Q_{i,1}$ and
$Q_{i,2}$ for the two job sizes respectively. When a job of size $p$ arrives, it goes to that machine $i$ for which $Q_{i,p}$ has the least load (load of a queue
is the total remaining processing time of jobs in it). Again, if the load on all such queues is more than $\alpha \opt$, we reject the job -- here $\alpha$ is 
a parameter which is $O(1/\eps)$. Thus, the algorithm 
ensures that all queue sizes remain bounded. The non-triviality lies in figuring out which job  a machine chooses to process.
Given the above dispatch rule, popular and simple heuristics like always processing  the shortest sized job or the job with  shortest remaining processing time 
will not guarantee bounded rejection ratio: 
Below we provide an example on how this fails.

Assume $\eps=1/4$, i.e., we are allowed to reject at most $\frac{1}{4}$th of the jobs; 
and the maximum queue length of any job size which
our online algorithm can afford on any machine is $4\opt$. In this example, $\opt$ 
will be $2$. Suppose there are 8 machines $m_1$, $m_2$, $\cdots$, $m_8$; also let the jobs can be of two sizes $1$
and $2$. At time $t=0$, seven jobs each of size $2$ and one of size $1$ arrive; the online algorithm sees the jobs in the following order: the first four jobs 
of size $2$ can be processed on any machine,
so our algorithm dispatches them to machines, say  $m_1$, $m_2$, $\cdots$, $m_4$;  the fifth (sixth) job of size $2$ can be processed on $m_1$, $m_2$ (respectively 
$m_3$, $m_4$), so our algorithm dispatches them to machines $m_1$ and $m_3$ respectively; the seventh job of size $2$ can be processed on $m_1$ only. The eighth job 
of size $1$ can be processed on any machine, suppose our algorithm dispatches it to $m_1$. 
Thus at end of time $t=0$, the load on machine $m_1$ for size $2$ jobs and size $1$ jobs are $6$ and $1$ respectively. 
Since our processing rule picks the shortest sized (or the shortest remaining processing timed) job, the $1$ sized job will be
processed on machine $m_1$ at this time step. Now suppose for every time steps $t=1,2,\cdots$, a job of size $1$ arrives which can 
be processed on any machine, and for every alternate time steps $t=2,4,6, \cdots$, a job of size $2$ arrives which can 
be processed only on machine $m_1$. Suppose our algorithm dispatches the size $1$ jobs to $m_1$ (since the algorithm will 
always prefer  processing size $1$ jobs over size $2$ jobs, the queue size of size $1$ jobs on all machines at the end of a time step 
will be zero, and hence the algorithm may dispatch the size $1$ job arriving in the next time step to $m_1$). The algorithm will start 
rejecting all the size $2$ jobs arriving in time $4,6,8,\cdots$. Thus the fraction of the rejected jobs will be close to $1/2$.
 
Note that the optimum solution of the scenario will always dispatch the size $2$ jobs of time steps $t=2,4,6, \cdots$ to  $m_1$ 
and the size $1$ jobs to some other machine. Thus the optimal solution will  have a maximum flow time of $2$.

The processing policy which our algorithm employs is as follows. For a machine $i$, 
let $\load_{i,1}(t)$ and $\load_{i,2}(t)$ denote the load in the two queues at time $t$. At time $t$, the machine processes job from the queue for
which $\load_{i,p}(t)/p$ is largest -- this quantity is roughly equal to the number of jobs in the corresponding queue. Let us see why this strategy works. 
First we consider jobs of size 1. Note that the queue for size 1 jobs can contain up to $\alpha \opt$ jobs, while that for size 2 jobs can only contain 
$\alpha \opt/2$ jobs. As in the above case, we can define the intervals $\intunit{i}{l}$ for the size 1 queues, and write down inequalities~(\ref{eqf:example})
as long as $l \geq \alpha/2$ (because as long as the load on size 1 queue is more than $\alpha \opt/2$, the machine will not give preference to a size 2 
job, and so we can pretend that there are size 1 jobs only). This suffices to bound the number of size 1 jobs which get rejected in terms of the total number
of size 1 jobs that has arrived.

When we need to bound the number of size 2 rejected jobs, we need to define the intervals more carefully. For a parameter $l$, $\intunit{i}{l}$ consists
of those minimal intervals $I$ where the load of $Q_{i,1}$ is at most $(l-1) \opt/2$ {\em and} that  of $Q_{i,2}$ is at most $(l-1) \opt$ at the left 
end point of $I$, whereas the load of $Q_{i,1}$  is at least $l \opt/2$ {\em or} that of $Q_{i,2}$ is at least $l \opt$ at the right end-point. Using these
intervals, one can again write down inequalities similar to~(\ref{eqf:example}) involving both size 1 and size 2 jobs. Using these inequalities, one can 
prove that the total number of rejected size 2 jobs is at most the total number of size 2 jobs and {\em half} the total number of size 1 jobs. 

More generally, suppose jobs sizes are powers of 2. Then each machine $i$ maintains a separate queue  $Q_{i,k}$ for jobs of size $2^k$, and jobs are dispatched 
to machines according to load on corresponding queues only. At any time $t$, a machine $i$ processes job from the  queue $Q_{i,k}$ 
for  which the load divided by 
$2^k$ is highest. Another way of thinking about this rule, which forms the intuition in more general case, is that we prioritize the queues on a machine based
on how full they are. Suppose the queue $Q_{i,k}$ is full to a fraction of $f_k$ (i.e., load on it is equal to $f_k \cdot \alpha \opt$). Then, we
 prefer the queue for which $f_k/2^k$ is largest.

\noindent
{\bf Arbitrary weight and unit size} 
Again assume that all weights are powers of 2. Each machine $i$ maintains queues $Q_{i,w}$ for jobs with weight $2^w$, and as before, jobs are dispatched
based on their corresponding queue. A machine $i$ prioritizes these queues in the decreasing order of $w$. Drawing analogy from the above case,
 if $f_{w}$ denotes the 
fraction to which the queue $Q_{i,w}$ is full (note that a queue for jobs of weight $2^w$ is full, if the total number of jobs in it is $\alpha \opt/2^w$, where
$\alpha$ is $O(1/\eps)$), then machine $i$ picks the next job from the queue for which $f_{w}\cdot 2^w$ is highest.

\noindent
{\bf General Case}
For simplicity, assume all jobs sizes and weights are powers of 2. For each machine $i$, we have queues $Q_{i,w,p}$ for jobs of size $2^p$ and 
weight $2^w$. However, it is unclear how to prioritize these queues because we can have two queues $Q_{i,w_1, p_1}$ and $Q_{i,w_2,p_2}$
for which $w_1 < w_2$ but $p_1 < p_2$. Suppose the queues $Q_{i,w_1,p_1}$ and $Q_{i,w_2,p_2}$ are  full to the extent of $f_1$ and $f_2$
respectively. We know fraction $f_1$ for $Q_{i,w_1,p_1}$ corresponds to fraction $f_1/2^{p_1-p_2}$ for $Q_{i,w_1,p_2}$ (using the argument above
for unit weight and arbitrary size), and the fraction $f_1/2^{p_1-p_2}$ for $Q_{i,w_1,p_2}$ corresponds to fraction $f_1 \cdot \frac{2^{w_1-w_2}}{2^{p_1-p_2}}$
for $Q_{i,w_2,p_2}$ (using the argument for unit size and arbitrary weight). Thus, machine $i$ prefers a job from $Q_{i,w_2,p_2}$ over $Q_{i,w_1,p_1}$
iff $f_2 \geq f_1 \cdot \frac{2^{w_1-w_2}}{2^{p_1-p_2}}$. Our algorithm is based on this rule.

The proof that all queues remain bounded again relies on carefully defining a set of intervals, and corresponding dual variables for an LP relaxation. However 
there are subtle and non-trivial details. While defining the intervals for a particular type of jobs (say of size $2^p$ and weight $2^w$), it may happen that
the algorithm processes other kinds of jobs during these intervals: (i) if these jobs are of very high density, then we are completing lot of weight in small time, and
so this should be somehow beneficial, (ii) if there are jobs whose density is close to the density of such jobs, then we cannot handle these directly; in fact, we need
to go back and define these intervals for a {\em group} of jobs of similar density, and (iii) if these jobs have low density, then we need to somehow prove that
either they have very high weight and so need to finish soon in any solution, or else such jobs cannot be processed during such intervals. 

This completes our informal description of the first part of our algorithm where we ensure that queues remain of bounded size. The second part of the algorithm
makes sure that a job of weight $w_j$ does not starve for too long (much longer than a constant times $\opt/w_j$). To accomplish this, it runs the algorithm above
in background. If the first algorithm tries to process a job $j$ at some time $t$ on a machine, the second algorithm processes $j$ as well unless there is a job
of much higher density in the queue of machine $i$. If the latter case happens, it processes such a job. A job which waits for too long (compared to $\opt/w_j$)
gets rejected. We prove that the weight of such jobs is small. The difficulty in the analysis arises for the following reasons: (i) the algorithm $\A$ ensures
that the size of each of the queues on a machine remains within a limit, whereas we would like the {\em total} size of the queues to remain close to this limit; to
ensure this, we need to prove that many of these jobs will get rejected, (ii)
we may reject a job after processing
it for a while -- the time used for processing such jobs gets wasted, and so we need to ensure that this remains very small.

Finally, we consider the more general problem {\wtdlinfgen} where a job $j$ has two weights -- $w_j^f$ and $w_j^r$. Recall that the weight $w_j^f$ is used for computing weighted 
flow-time, whereas we do not want to reject jobs whose total $w_j^r$ weight is high.
This is a generalised version of the problem of  minimizing the maximum stretch.
It turns out that the first part of the algorithm described above carries over
without much change to this problem as well. However, the second part of the algorithm requires some subtle changes -- in particular, we cannot bound
the amount of time in which a machine processes jobs which eventually get rejected, and so the above ideas do not apply directly.

\section{Algorithm for the {\loadbal} problem}
\label{secf:loadbal}
In this section, we describe our algorithm for the {\loadbal} problem. We first consider a special case when all jobs have size 1. This will illustrate the main ideas involved, and then we shall extend it to the case of general processing time. We shall assume that we know the optimal value, denoted by $\opt$. We shall show how to get rid of this assumption later.

\subsection{Unit size jobs}
Our  algorithm $\A$ is greedy. We pick a threshold $\alpha$ (which depends on $\eps$) to be defined later. Each
machine maintains the current {\em load} on it, i.e., the number of jobs assigned to it so far (not counting the jobs which have been rejected).
Let $\load_i(t)$ be the load on machine $i$ at time $t$.
When a
job $j$ arrives, it is dispatched to the machine in $S_j$ with the least load, provided the load on this machine is less
than $\alpha \opt$. If the load on this machine happens to be greater than $\alpha \opt$, we reject this job.

Clearly, our algorithm ensures that we do not exceed the load on a machine by more than $\alpha \opt$. We need
to prove that we will not reject more than $\eps$ fraction of the jobs.
 Suppose this fact is not true, and let
$n$ be the first time when we reject more than $\eps$ fraction of the jobs --
call these rejected jobs $\jrej$.

 We divide the machines into several groups. Let
$M_\alpha$ denote the set of machines $i$ for which $\load_i(n) = \alpha \opt$. For an integer $s$, $0 \leq s < \alpha$,
let $M_s$ be the set of machines $i$ for which $\load_i(n) \geq s \cdot \opt$. Note that $M_s \subseteq M_{s-1}$.
 Let $m_s$ denote $|M_s|$.
Now we show that $m_s$ increases exponentially as we decrease $s$.

Consider a job $j$ which arrives at time $t$. Suppose it is dispatched to machine $i$. Define $\level(j)$
as $\load_i(t)$ (not counting  $j$) -- we can think of this as the position of $j$ in the queue of machine $i$. For an index
$u$, let $J(u)$ be the set of jobs whose level is at least  $u$.

\begin{claim}
\label{clf:main}
For any $s$, $0 \leq s < \alpha$, $m_{s} \geq 2^{\alpha-1-s} \cdot m_\alpha$.
\end{claim}
\begin{proof}
For any job $j \in J(s \opt),$ $S_j \subseteq M_s$. Indeed,  if a machine  $i \notin M_{s} $, then
$\load_i(n) < s \opt$. Further if  $i \in S_j$, then $j$ had the option of getting dispatched to $i$, and then
$\level(j)$ should be less than  $s \opt$, a contradiction. Therefore, we get the following lower bound on $\opt$:
\begin{eqnarray}
\label{eqf:1}
 \opt \geq \frac{|J(s \opt)|}{m_{s}}.
 \end{eqnarray}
Now, it is easy to see that $|J(s\opt)|$ is at least $\opt \cdot (m_{s+1} +  \ldots + m_{\alpha})$. Substituting
this in the lower bound above, we get $m_s \geq m_{s+1} + \ldots + m_{\alpha}$.
\end{proof}

\begin{corollary}
\label{corf:main}
The number of jobs is at least $(2^\alpha-1) \cdot \opt \cdot m_\alpha.$
\end{corollary}
\begin{proof}
Using the lemma above, the number of jobs is at least
$$ (m_0 + m_1 + \ldots + m_{\alpha-1}) \opt \geq (2^\alpha-1) \cdot m_\alpha \opt. $$.
\end{proof}

Corollary~\ref{corf:main} implies that $|\jrej|$ is more than  $\eps \cdot (2^\alpha-1) \cdot \opt m_0 > \opt m_0$, if we pick
$\alpha$ to be $\log \left( \frac{1}{\eps} \right)+2$ .
Now, observe that for any job $j \in \jrej$, $S_j \subseteq M_0$. Indeed, the fact this job is rejected means
that all the machines in $S_j$ were loaded to the maximum capacity. Now we get a contradiction -- we have
a set of jobs $\jrej$ of cardinality greater than $\opt m_\alpha$ which can be scheduled only on a subset of $m_\alpha$ machines. Hence, the optimal value must be larger than $\opt$.

\noindent
{\bf Remarks:} If instead of unit size jobs, the processing times of jobs were in the range $[1,2]$, the above analysis would still apply provided we pick $\alpha$ to be slightly larger -- $2 \cdot \log \left( \frac{1}{\eps} \right)+2$. The algorithm remains unchanged, except for
the fact that $\load_i(t)$ is now defined as the total processing times of jobs dispatched to machine $i$ till time $t$ .
We highlight the main changes in the analysis. Claim~\ref{clf:main} continues to hold -- in inequality~(\ref{eqf:1}), the numerator on
the right hand side gets replaced by the total processing time of jobs in $J(s \opt)$. Corollary~\ref{corf:main} holds with a
weaker bound of $(2^{\alpha-1}-1) \cdot \opt m_\alpha$ on the number of jobs -- this is because the jobs could have size 2. Rest of the arguments remain
unchanged. Thus, we get the following result.

\begin{theorem}
\label{thmf:unitjobload}
Consider an input instance where  all job sizes lie between 1 and 2, and the optimum value is  $\opt$. Then the algorithm described above is $O(\alpha)$-competitive and rejects at most $\eps$-fraction of the jobs.
\end{theorem}

\subsection{General Processing Times}
\label{secf:gen}

We now extend the above result to general processing times. Some definitions first. We say that a job $j$ is of class $k$ if
$p_j \in [2^k, 2^{k+1})$. We do not know the smallest job size, and so cannot assume (by scaling) that all job sizes are at least 1.
Hence, the class of a job could be negative as well. We say that a set of jobs is {\em $\Delta$-separated} for a positive integer
$\Delta > 0$, if there exists an integer $k$ such that the class of any job $j$ in this set belongs to $\{k + i \Delta: i \mbox{ is an integer} \}$.

\noindent
{\bf $\Delta$-separated jobs:}
We first assume that the jobs are $\Delta$-separated. Our scheduling algorithm works in two stages. In the first  stage, it may violate the
objective value by a  large amount, but this will get fixed in the second stage. However, we shall ensure that in the first stage, the
number of rejected jobs is at most $\eps$-fraction of the total number of jobs. As before, assume that we know the optimal value $\opt$.

\noindent
{\bf Stage 1:} Each machine $i$ maintains a queue of jobs assigned to it. For a class $l$, let $\load_{i,l}(t)$ denote the total processing time
of jobs of class  $l$ in the queue of machine $i$ at time $t$. We say that $i$ is {\em full with respect to class $l$} at time $t$  if $\load_{i,l}(t) \geq \alpha \cdot \opt,$
for some parameter $\alpha$ to be specified later. The dispatch algorithm is as follows: when a job $j$ of class $l$ arrives at time $t$, dispatch
it to the machine $i \in S(j)$ for which $\load_{i,l}(t)$ is smallest; unless all such machines are full with respect to class $l$. If the latter
case happens, we reject the job.

\noindent
{\bf Stage 2:} This is a pruning step. Note that (and this is important) this step does not affect stage~1 at all. So, while computing
$\load_{i,l}(t)$
in the algorithm described above, we will assume that no pruning happens. However, without this step, the queue sizes may go much beyond
$\opt$. For a machine $i$, and time $t$, let $\load_i(t) = \sum_l \load_{i,l}(t)$ denote the total load on this machine at time $t$. In this
stage, we do the following for each machine $i$ and time $t$: if at time $t$, $\load_i(t) > 2 \alpha \cdot \opt$, we keep removing the largest
jobs in the queue of machine $i$ till $\load_i(t)$ becomes at most $2 \alpha \cdot \opt$ (we can assume a fixed way of breaking ties).

This completes the description of the algorithm. We now analyze the algorithm.
We fix a time $t^\star$ (and we assume that the offline optimum has objective value $\opt$).

For a fixed class $l$,
the stage 1 algorithm is same as the algorithm for unit size jobs (or when the job sizes differ by a factor of 2 only) described in the previous section. Hence, Theorem~\ref{thmf:unitjobload} implies that
the total processing time of jobs of class $l$ dispatched to any particular machine  is at most
$\alpha \cdot \opt$. Further, the total number of rejected jobs in stage 1 is only $\eps$-times the total number of jobs.

Now, in Stage 2, we ensure that the total load on a machine is at most $2 \alpha \cdot \opt$. We need to bound the number of jobs which
get rejected. Fix a machine $i$. Let the jobs which get dispatched to $i$ (in Stage 1) till $t^\star$ be $j_1, \ldots, j_n$. We assume that the total processing time of these jobs is more than $2 \alpha \opt$, otherwise we do not reject any of these jobs in Stage 2.
 Also assume that these jobs are arranged
in ascending order of processing time. Let $r_\opt$ be the smallest index $u$ for which $p_{j_1} + \cdots + p_{j_{u}} \geq \alpha \cdot \opt$ and $r_{2\opt}$ be the smallest index $u$ such that $p_{j_1} + \cdots + p_{j_{u}} \geq 2 \alpha \cdot \opt$. Clearly, $r_{2\opt} \geq r_\opt$.
We make some important observations:
\begin{itemize}
\item[(i)] During stage 2, we will not remove the jobs $j_1, \ldots, j_{r_{2\opt}}$. Indeed, in order to remove a job $j$, there must be at least $2 \cdot \alpha \cdot \opt$ volume of smaller jobs.
\item[(ii)] During stage 2, we will remove all jobs $j_u$ for $u > r_{2\opt}$. This is true because  at time $t^\star$, we have at least $2 \cdot \alpha \cdot \opt$ volume of jobs which are smaller than this job.
\item[(iii)] The class of job $j_{r_\opt}$ is strictly less than that of $j_{r_{2\opt}}$: Since $j_{r_\opt}$ comes before $j_{r_{2\opt}}$ in the ordering, the class of $j_{r_{\opt}}$ is at most that of $j_{r_{2\opt}}$. Suppose the two jobs are of the same class. Then, the jobs
    in $j_{r_\opt}, j_{r_\opt+1}, \ldots, j_{r_{2\opt}}$ are of the same class, and their total processing volume is strictly larger than $\alpha \opt$. But this is a contradiction -- we argued above that for any particular class, we will not dispatch more than $\alpha \cdot \opt$ volume to a machine. Since the jobs are $\Delta$-separated, class of $j_{r_\opt}$ is at least $\Delta$ less than that of $j_{r_{2\opt}}$.
\end{itemize}

Let $c_\opt$ and $c_{2\opt}$ denote the class of the jobs $j_{r_\opt}$ and $j_{r_{2\opt}}$ respectively. As observed in (iii) above, $c_{2\opt} \geq c_\opt + \Delta$.
We are only rejecting jobs of class $c_{2\opt}$ or higher (from the queue of machine $i$). Further, the total number of jobs of a class $l$ that
can be rejected is at most $\frac{\alpha \cdot \opt}{2^l}$, because the total volume of jobs of class $l$ assigned to this machine is at most
$\alpha \cdot \opt$. Thus, the total number of rejected jobs (among those dispatched to machine $i$) in Stage 2 is at most
$$\sum_{l \geq c_{2\opt}} \frac{\alpha \cdot \opt}{2^l} \leq \frac{\alpha \cdot \opt}{2^{c_{2\opt}-1}} \leq \frac{\alpha \cdot \opt}{2^{c_{\opt}+\Delta-1}}
\leq \frac{\eps \cdot \alpha \cdot \opt}{2^{c_{\opt}+1}}, $$
provided $\Delta=\log \left( \frac{1}{\eps} \right) + 2.$ Now, observe that the jobs in $j_1, \ldots, j_{r_\opt}$ have volume at least $\alpha \cdot \opt$, and their class
is at most $c_\opt$. Hence, the number of these jobs, $r_\opt$, is at least $$\frac{\alpha \cdot \opt}{2^{c_\opt+1}}.$$ Comparing this with the expression above, we see that at most $\eps$ fraction of the jobs are removed in a Stage 2.

Combining the above observations, we get (replacing $\eps$ by $\eps/2$ in the argument above)
\begin{lemma}
\label{lemf:sep}
Assuming we know the value of the offline optimum  and that the  jobs are $\Delta$-separated, the above algorithm is $O \left( \log \left( \frac{1}{\eps} \right)\right)$-competitive,
where $\Delta=\log \left( \frac{1}{\eps} \right) + 2.$ Further, it rejects at most $\eps$-fraction of the jobs.
\end{lemma}

Any set of jobs can be partitioned into $\Delta$ disjoint sets, each of which is $\Delta$-separated. Running the above algorithm on each such set independently, we get
\begin{corollary}
\label{corf:loadbal}
Assuming  the value of the offline optimum is at most $\opt$, the algorithm described above  assigns load  $O \left( \log^2 \left( \frac{1}{\eps} \right)\right) \cdot \opt$ on any machine, and rejects at
most $\eps$-fraction of the jobs.
\end{corollary}

\subsection{Removing the assumption about $\opt$}
\label{secf:remload}

So far we have assumed that we know the value of the offline optimum solution. We now show how to relax this assumption. The idea is fairly standard. We start with a small guess for $\opt$, and double it whenever we realize that the guess was less than the actual offline optimum value. The details are given in Figure~\ref{figf:load}. For a parameter $\opt$, let $\A(\opt)$ denote the algorithm given by Corollary~\ref{corf:loadbal} where the estimate for the optimal value is $\opt$. We shall call each iteration of {\tt Step~2} as a {\em phase}. Note that when we run $\A(\opt)$ in the beginning of a phase, the algorithm ignores the jobs that have been assigned in the previous phase. It only considers the jobs that arrive next (and hence, quantities like $\load_i(t)$ needed by $\A(\opt)$ do not take into account the jobs which were assigned in previous phases).

\begin{figure}[ht]
   \begin{center}
     \begin{boxedminipage}{6.5in}
          {\bf Load Balance:} \medskip\\
        \sp \sp 1.   Initialize $\opt \leftarrow p_j,$ where $j$ is the first released job. \\
         \sp \sp 2. {\bf Repeat} \\
         \sp \sp \sp \sp (i) Run $\A(\opt)$  on the jobs which arrive next.  \\
         \sp \sp \sp \sp (ii) If the algorithm rejects more than $\eps$-fraction of the jobs which  arrived \\
         \sp \sp \sp \sp \sp \sp \sp \sp after the time it started running, \\
         \sp \sp \sp \sp \sp \sp \sp \sp \sp \sp  Stop the algorithm and update
         $\opt \leftarrow 2 \opt$.
      \end{boxedminipage}
       \caption{Algorithm for {\loadbal} }
       \label{figf:load}
       \end{center}
 \end{figure}

Now we analyze the competitive ratio of this algorithm. It is clear that it does not reject more than $\eps$-fraction of the jobs because in each phase, the algorithm rejects at most $\eps$-fraction of the jobs which arrive in that phase. Let $T^\O$ denote the objective value of the offline optimum solution. We first argue that the estimate $\opt$ always stays (almost) below $T^\O$.
\begin{claim}
\label{clf:guess}
Except perhaps for the last phase, the value of $\opt$ is at most $T^\O$.
\end{claim}
\begin{proof}
Consider the first phase for which  $\opt$ is more than $T^\O$.
Note that the optimal value for the set of jobs which arrive during this period will be at most $T^\O$, which is at most $\opt$. Corollary~\ref{corf:loadbal} now implies that $\A(\opt)$ will not reject more than $\eps$-fraction of the jobs released during this phase, and so this is the last phase of the algorithm.
\end{proof}

The main result  now follows easily from the above claim.
\begin{theorem}
\label{thmf:mainloadbal}
The algorithm {\tt Load Balance} is $O\left( \log^2 \left( \frac{1}{\eps} \right)\right)$-competitive and rejects at most $\eps$-fraction of the jobs.
\end{theorem}
\begin{proof}
We have already argued that for any time $t$, the total number of jobs rejected by the algorithm till time $t$ is at most $\eps$-fraction of the jobs released till this time. The final value of $\opt$ is at most $2 T^\O$. Since $\opt$ increases by a factor of 2 after each phase, Corollary~\ref{corf:loadbal} and Claim~\ref{clf:guess} imply that the total load assigned to a particular machine is at most
$$ O \left( \log^2 \left( \frac{1}{\eps} \right)\right) \cdot
\left( 2 T^\O + \frac{2 T^\O}{2} + \frac{2 T^\O}{4} + \ldots \right)
\leq O \left( \log^2 \left( \frac{1}{\eps} \right)\right) T^\O.$$
This proves the desired result.
\end{proof}

\section{Algorithms for the {\linf} problem}
\label{secf:linf}
In this section, we consider the {\linf} problem. Again, we shall begin by assuming that we know the optimal value $\opt$.
In Section~\ref{secf:unitsize}, we consider the special case when all jobs have unit size. Here,
the algorithm turns out to be a natural one: when a job arrives, send it to the least loaded machine (among the ones it can be processed on).
However, the extension to arbitrary job sizes and arbitrary weight turns out to be quite tricky, and although the algorithm remains simple, it does not correspond to a natural idea. We give details of this algorithm in Section~\ref{secf:gensize}.
Finally, we show how to remove the assumption about the knowledge of $\opt$.

\subsection{Unit Job Size}
\label{secf:unitsize}
We consider the special case when all jobs have size 1.
The algorithm is  greedy: for each machine $i$ and time $t$, it maintains the
number of jobs waiting in the queue of machine  $i$ at (the beginning of) time $t$,
call this value $\load_i(t)$. When a new job $j$ arrives at time $t$,
it gets dispatched to the machine $i \in S_j$ for which $\load_i(t)$ is smallest, unless the load on every machine
$i \in S_j$ is at least $\alpha \opt,$ where $\alpha = \frac{1}{\eps}$.
If the latter happens, we reject the job. Each machines processes jobs in order they are dispatched to it.
This completes the description of the algorithm.

\noindent
{\bf Analysis:}
 We first give a way of proving lower bounds on the optimum value.
A {\em machine interval}  is defined as  a pair $(I,i)$ where $I$ is an interval and $i$ is a machine.
\begin{lemma}
\label{lemf:lpdual}
Let $\bI$ be a set of machine intervals and $J$ be a set of jobs. Let $\alpha_j$ be non-negative values assigned to the
 jobs $j \in J$ such that the following  condition is satisfied for all jobs $j$ and machines $i \in S_j$:
$$\alpha_j \leq | \{(I,i) \in \bI :  r_j \in I\} |.$$
Then,
\begin{eqnarray}
\label{eqf:lpdual}
\opt \geq \frac{\sum_j \alpha_j - \sum_{(I,i) \in \bI} \length(I)}{|\bI|}
\end{eqnarray}
\end{lemma}

\begin{proof}
This result follows from weak duality. Consider the following LP relaxation for (the offline optimum of) this problem.
For a job $j$ and machine $i \in S_j$, we have a variable $x_{ij}$ which is 1 iff
$j$ is dispatched to $i$. It is easy to check that the following is a valid LP relaxation:
\begin{align*}
\min & \ \ \ \  T \\
T & \geq \sum_{j: r_j \in I, i \in S_j} x_{ij} - \length(I) \ \ \ \ \ \mbox{for all machine intervals $(I,i)$} \\
\sum_{i : i \in S_j} x_{ij} & \geq   1 \ \ \ \ \ \  \ \mbox{for all jobs $j$} \\
x_{ij} & \geq 0 \ \ \ \ \ \ \mbox{for all $i,j$}
\end{align*}

The dual LP is as follows (variables are $\alpha_j$ and $\beta_{(I,i)}$):
\begin{align*}
\max \ \ \ \ \ \ \ \ &  \sum_j \alpha_j  - \sum_{(I,i)} \length(I) \cdot \beta_{(I,i)}\\
\alpha_j & \leq \sum_{(I,i): r_j \in I, i \in S_j} \beta_{(I,i)} \ \ \ \ \ \mbox{for all $j$ and $i\in S_j$} \\
\sum_{(I,i)} \beta_{(I,i)} & \leq 1 \\
\alpha_j, \beta_{(I,i)} & \geq 0
\end{align*}

The desired lemma now follows by setting $\beta_{(I,i)} = 1$ for all machine intervals $(I,i) \in \bI$.
\end{proof}

We now prove correctness of our algorithm. It is easy to see that the flow-time of any job which does not get rejected is
at most $\alpha \opt$. When a job is dispatched to a machine, the load on the machine is at most $\alpha \cdot \opt$, and since
the machine processes jobs in the order in which they get dispatched to it, this job will finish within $\alpha \cdot \opt$ time.
We now need to bound the number of rejected jobs. Fix a time $t^\star$ -- we shall bound the number of rejected jobs till time $t^\star$.
Recall that we are assuming that the value of the optimal solution for this input (till time $t^\star$) is at most $\opt$.

We now give some definitions. For a time $t$ and machine interval $(I,i)$, we shall often abuse notation and say $t$ {\em lies in} $(I,i)$ (or
$(I,i)$ {\em contains} $t$) when
$t \in I$. Similarly, we shall say that two machine intervals $(I_1, i_1)$ and $(I_2, i_2)$ are {\em disjoint} if the corresponding intervals
$I_1$ and $I_2$ are disjoint.
 Given a set $\bI$ of machine intervals, the {\em set of intervals} in $\bI$ refers to the multi-set $\{I: (I,i) \in \bI\}$.
 For each index $l$, $0 \leq l
< \alpha$, and machine $i$, we define a set (possibly empty) of mutually disjoint machine intervals
$\intunit{i}{l}$.
The procedure for defining these intervals is described in Figure~\ref{figf:int}.

 \begin{figure}[ht]
   \begin{center}
     \begin{boxedminipage}{6.5in}
          {\bf Constructing $\intunit{i}{l}$:} \medskip\\
        \sp \sp 1.   Initialize $t \leftarrow t^\star$. \\
         \sp \sp 2. {\bf Repeat} \\
         \sp \sp \sp \sp (i) Let $t_2$ be the highest  time before (or equal to) $t$ such that $\load_i(t_2) \geq  l \cdot \opt.$ \\
         \sp \sp \sp \sp (ii) Let $t_1$ be the earliest time before $t_2$ such that $\load_i(t') >   (l-1) \opt$ for all
                              $t' \in (t_1, t_2]$. \\
         \sp \sp \sp \sp (iii) Add the machine interval $([t_1, t_2],i)$ to $\intunit{i}{l}$. \\
         \sp \sp \sp \sp (iv) $t \leftarrow t_1$. \\
         \sp \sp 3. {\bf Until} No more machine intervals can be added to $\intunit{i}{l}$.
      \end{boxedminipage}
       \caption{Construction of the set of intervals in $\intunit{i}{l}$}
       \label{figf:int}
       \end{center}
 \end{figure}

It is easy to see that  for any value of $l$ and machine $i$, the set of machine intervals in $\intunit{i}{l}$ are mutually disjoint.
We now prove a few simple properties of these sets of intervals. For a job $j$ which gets released at time $t$
and dispatched to machine $i$, let $\level(j)$ denote $\load_i(t)$ (without counting the job $j$ itself).
\begin{lemma}
\label{lemf:prop}
Fix a machine $i$ and parameter $l$. The sets of machine intervals $\intunit{i}{l}$ satisfy the following properties:
\begin{itemize}
\item {\em (covering)} Suppose a job $j$ satisfies $\level(j) \geq  l \cdot \opt$, and assume $i \in S_j$.
Then, there exists a machine interval $(I,i) \in \intunit{i}{l}$
such that $r_j \in I$.
\item {\em (nesting)} Given any machine interval $(I_l,i) \in \intunit{i}{l}$, there exists a machine interval $(I_{l-1},i) \in \intunit{i}{l-1}$
such that $I_{l-1}$ contains $I_l$.
\item For any machine interval $(I_l,i) \in \intunit{i}{l}$, the total number of jobs which get dispatched to $i$ during $I_l$ is at
least $\length(I_l) + \opt$.
\end{itemize}
\end{lemma}
\begin{proof}
Consider a job $j$ such that  $\level(j) \geq  l \cdot  \opt$, and a machine $i \in S_j$. Then $\load_i(r_j) \geq  l \cdot \opt$ as
well.
Suppose, for the sake of contradiction that there
is no machine interval in $\intunit{i}{l}$ such that $r_j$ lies in it. Among the set of intervals in $\intunit{i}{l}$, let
 $I_f$ be the first interval (if any)
to the right of $r_j$, and suppose $s$ is the starting time of $I_f$ (if $I_f$ does not exist, $s$ is defined
as $t^\star$). After adding $(I_f,i)$ to $\intunit{i}{l}$, our algorithm for constructing $\intunit{i}{l}$ would have tried
$t_2=r_j$, and hence there should be a machine interval  containing $r_j$ in $\intunit{i}{l}$.

The argument above shows that if for any time $t$, $\load_i(t) \geq  l \cdot \opt$,
then there exists a machine interval in
$(I,i) \in \intunit{i}{l}$ containing $t$. Consider  $(I_l=(s_l,e_l),i) \in \intunit{i}{l}$. Since $\load_i(e_l) \geq
 l \opt,$ there exists an interval $(I_{l-1}=(s_{l-1},e_{l-1}),i) \in \intunit{i}{l-1}$ such that
$e_l \in I_{l-1}$. Since $\load_i(t) > (l-1) \opt $ for all $t \in (s_l, e_l)$, the left end-point of
$I_{l-1}$ will appear before $s_l$.

The third claim is easy to see. If $I_l = [s,t]$, then $\load_i(t)-\load_i(s) \geq \opt$. Further, the
machine is busy during $[s,t]$, and so must have processed $t-s$ volume during this period.
\end{proof}

For an index $l$, let $\ints{l}$
denote the union over all machines $i$ of the machine intervals in $\intunit{i}{l}$.  Let
$\len{l}$ denote the total length of the intervals in $\ints{l}$, and $m_l$ denote the cardinality of $\ints{l}$.
 Let $\jobs{l}$ be the set of jobs which get dispatched to  a machine $i$ during the intervals in $\intunit{i}{l}$.


Let $\bI$ denote $\cup_{l=1}^{\alpha} \ints{l}$. Let $\jrej$ denote the set of rejected jobs.
Consider the following assignment of dual variables: if $j \in \jrej$, then we set $\alpha_j = \alpha$, otherwise
if $\level(j) \in [(l-1) \cdot \opt, l \opt)$, then we set $\alpha_j = l-1$. The following claim is easy to see.
\begin{claim}
\label{clf:feas}
The set of machine intervals $\bI$ and the $\alpha_j$ values defined above satisfy the conditions of Lemma~\ref{lemf:lpdual}.
\end{claim}
\begin{proof}
If $j \in \jrej$, $\load_i(r_j) \geq \alpha \cdot \opt$ for all $i \in S_j$.
Lemma~\ref{lemf:prop}
now shows that for any $i \in S_j$ and all $l, 1 \leq l \leq \alpha,$ there exists a machine interval $(I,i) \in \intunit{i}{l}$ containing $r_j$. Hence,
for any $i \in S_j$,
$|\{(I,i): r_j \in I\}|$ is equal to $\alpha$. Now suppose $j \notin \jrej$. Let $k$ denote $\alpha_j$. Since $\level(j) \geq k \cdot \opt$, Lemma~\ref{lemf:prop}
again implies that if $i \in S_j$, then for all $l $, $1 \leq l \leq k$, there is a machine interval $(I,i) \in \intunit{i}{l}$ such that $r_j \in I$. Hence, the desired result holds in this case as well.
\end{proof}

Using the above claim, we can applying Lemma~\ref{lemf:lpdual} to $\bI$ and $\alpha_j$ values to get
\begin{eqnarray}
\label{eqf:key}
\opt \cdot (m_1 + \ldots + m_\alpha) \geq \sum_j \alpha_j - \sum_{l=1}^\alpha \len{l}
\end{eqnarray}

\begin{claim}
\label{clf:nonrej}
$$\sum_{j \notin \jrej} \alpha_j \geq \sum_{l > 1} \len{l} + \sum_{l > 1} m_l \opt.$$
\end{claim}
 \begin{proof}
 We first argue that $\sum_{j \notin \jrej} \alpha_j \geq \sum_{l > 1} |\jobs{l}|$. Consider a job $j$ with $\alpha_j = k$. This implies that
 $\level(j) < (k+1) \cdot \opt$. So $j \notin \jobs{l}$ for any $l > k+1$. Indeed, suppose $j$ is dispatched to machine $i$ and there is
 a machine interval $(I,i) \in \intunit{i}{l}$ for some $l > k+1$ such that $r_j \in I$.
 Then the load on machine $i$ stays above $(k+1) \cdot \opt$ during $I$. So $\level(j) \geq (k+1) \cdot \opt$, a contradiction. Therefore,
 $j$ can only contribute towards $\jobs{2}, \ldots, \jobs{k+1}$ in the right hand side of the above inequality. Hence,
 $\sum_{j \notin \jrej} \alpha_j \geq \sum_{l > 1} |\jobs{l}|$.
The claim now follows from Lemma~\ref{lemf:prop} (third part).
 \end{proof}

We can now show that the number of rejected jobs is small.
\begin{corollary}
\label{corf:unitsizerejected}
The total number of rejected jobs is at most $\eps$ times the total number of jobs. Further, the flow-time of
any job is at most $\opt/\eps$.
\end{corollary}
\begin{proof}
Applying Claim~\ref{clf:nonrej} to inequality~(\ref{eqf:key}), we get
$$ \opt \cdot m_1 + \len{1} \geq \sum_{j \in \jrej} \alpha_j = \alpha \cdot |\jrej| . $$
Observe that the number of jobs is at least $\opt \cdot m_1 + \len{1}$ -- this follows from the
third part of Lemma~\ref{lemf:prop} and the fact that for any machine $i$, all intervals in $\intunit{i}{1}$ are disjoint.
If we pick $\alpha=\frac{1}{\eps}$, the desired result follows. The second statement follows from the fact that
the queue size never exceeds $\alpha \opt$.
\end{proof}

\subsection{Extension to {\wtdlinf}}
\label{secf:gensize}

We now extend the above result to the case of arbitrary job sizes and associated weights. We can assume without loss of
generality that weight of a job is a power of 2 -- this will affect the objective function by a factor of 2 only. Further,
the weight of rejected jobs could be off by a factor of 2 -- but we can replace $\eps$ by $\eps/2$ to take care of this
(this will affect the flow-time of a job only by a constant factor).

We shall describe the scheduling algorithm in two stages. First we describe an algorithm $\A$ -- this algorithm may not
ensure that flow-time of all jobs (which are not rejected) are small, but it will ensure that the queue sizes at all times
(on any machine) will be small. Our actual algorithm, which we call $\B$, will use $\A$ in the {\em background}. It will
emulate $\A$, but will periodically reject more jobs, and may prefer to process higher density jobs at a time slot (as compared
to $\A$).

\vspace*{1mm}
\noindent
{\bf Algorithm {$\A$}: }\
Before we describe the algorithm $\A$, we give some definitions. These definitions group jobs into various {\em classes}
depending on their processing requirement or weight or density. We say that a job $j$ is of {\em size class} $p$ if
$p_j \in [2^p, 2^{p+1})$. We say that it is of {\em weight class} $w$ if $w_j = 2^w$ (recall that weight of a job is
a power of 2), and of density class $d$ if its density, i.e., $w_j/p_j$ lies in the range $[2^d, 2^{d+1})$. Observe that if
the size, weight and density classes of a job are $p,w,$ and $d$ respectively, then $d=w-p-1$. 

 Note that we cannot assume any lower bound on $p_j$ or $w_j$, and so, the classes could be negative.
Define $\type(j) = (w,d)$, if its density class is $d$, and its weight class is $w$.

Each machine $i$ maintains a queue for jobs of a particular type. For a machine $i$, time $t$, and pair $(w,d)$, let
$Q_{i,w,d}(t)$ denote the jobs of type $(w,d)$ waiting in the queue of machine $i$ at time $t$. Let
$\load_{i,w, d}(t)$ denote the total weighted remaining processing time of the jobs in $Q_{i,w,d}(t)$ -- if a job $j$ has remaining
processing time $p_j'$, then its {\em  weighted remaining processing time} is defined as $w_j p_j'$.

When a job $j$ of type $(w, d)$ arrives at time $t$, we dispatch it to the machine $i \in S_j$ for which $\load_{i,w,d}(t)$ is minimum, unless for all $i \in S_j,$  $\load_{i,w,d}(t) + p_j.2^{w} \geq \alpha^2 \cdot \opt$. If the latter case happens,  we reject this job. Here $\alpha = \frac{76}{\eps}$.
It remains to specify which job is processed at any time by a machine.

For a time $t$ and machine $i$, let $(w^\star_i(t), d^\star_i(t))$ be the pair $(w,d)$ with the highest $2^d \cdot \load_{i,w,d}(t)$ value.
  We process the earliest released job from the queue $Q_{i,w^\star_{i}(t), d^\star_{i}(t)}(t)$ on machine $i$ at time $t$. Assume a fixed rule of breaking ties.

This completes the description of the algorithm $\A$. We note a few important aspects: (i) For a machine $i$ and pair $(w,d)$, the algorithm always prefers the earliest released job in
the queue of type $(w,d)$ jobs. Hence, at any time $t$, there will be at most one job in $Q_{i,w,d}(t)$ which is partially processed, (ii) The policy which decides which job to process at a
time $t$ on a machine $i$ balances two aspects: it prefers jobs of higher density, but also prefers jobs for which the corresponding queue is close to the maximum limit -- the total 
weighted remaining processing times  of jobs in the queue $Q_{i,w,d}(t)$ should not exceed a constant times $\opt$.

\vspace*{1mm}
\noindent
{\bf Algorithm {$\B$}: }\ Now we describe the actual scheduling algorithm $\B$. When a job $j$ arrives at time $t$, it is dispatched according to $\A$: if $\A$ rejects this job, $\B$ also rejects
it; and if $\A$ dispatched it to machine $i$, then $\B$ also dispatches this job to $i$. Now, we describe the processing policy for a fixed machine $i$. Consider a time $t$. Let $d^\A(t)$
denote the density class of the job processed by $\A$ at time $t$. Then, $\B$ processes the following job at time $t$: if there is a job of density class at least $d^\A(t) + \logeps$ in the queue of
machine $i$ at time $t$, then $\B$ processes any such job; otherwise it processes the job of density class  $d^\A(t)$ with the highest weight class (it prefers the earliest released job in case of ties).
 Also note that if the second
case happens and there is no job of density $d^\A(t)$ in the queue of machine $i$ at time $t$ (in $\B$), we can process any job at this time.

The algorithm $\B$ may reject some more jobs. For a weight class $w$, we divide the time line into segments of length $\frac{\alpha^2 \opt}{\eps^2 2^w}$.
Suppose a job of weight class $w$ gets released during such a segment $S$, and let $S'$ be the segment immediately to the right of $S$. If
the job does not complete processing by the end of $S'$, the algorithm $\B$ rejects the job. This completes the description of $\B$.

\subsubsection{Analysis}
\label{secf:analysis}
We now analyze the scheduling algorithm. We first consider the algorithm $\A$ and prove that the queues remain bounded in size.

\noindent
{\bf Analysis for algorithm $\A$:}
 As in the case of unit sized jobs, we first give a lower
bound for $\opt$. A {\em weighted machine interval} is defined as a triplet $(I, i, w)$ where $I$ is a time interval, $i$ is a machine, and $w$ denotes a weight class. We say that a time
$t$ lies in (or belongs to) a weighted machine interval $(I,i,w)$ if $t \in I$. Similarly, we say that two machine intervals are disjoint (or nested) if the same holds for the associated time intervals.

\begin{lemma}
\label{lemf:genlower}
Let $\bI$ be a set of weighted machine intervals. Let $\alpha_j$ be non-negative values assigned to the jobs $j \in J$ such that the following  condition is satisfied for all jobs
$j$ and machines $i \in S_j$:
\begin{equation}
\label{eqf:dualcond}
\frac{\alpha_j}{p_j} \leq | \{(I,i,w') \in \bI:  r_j \in I, w' \leq w\} |,
\end{equation}
where $w$ denotes the weight class of $j$, i.e., $w_j = 2^w$.
Then,
\begin{eqnarray}
\label{eqf:lpdualgen}
\opt \geq \frac{\sum_j \alpha_j - \sum_{(I,i,w)\in \bI} \length(I)}{\sum_{(I,i,w) \in \bI}\frac{1}{2^w}}
\end{eqnarray}
\end{lemma}
\begin{proof}
The proof again follows from LP duality. It is easy to check that the following is a valid LP relaxation --
here $x_{ij}$ is a variable which is 1 if job $j$ is assigned to machine $i$, 0 otherwise.
\begin{align*}
\min & \ \ T \\
\sum_{j: r_j \in I, i \in S_j, w_j\geq 2^w} p_{j}\cdot x_{ij} - \length(I) & \leq \frac{T}{2^w} \ \ \ \ \ \mbox{for all weighted machine intervals $(I,i,w)$} \\
\sum_{i : i \in S_j} x_{ij} & \geq   1 \ \ \ \ \ \  \ \mbox{for all jobs $j$} \\
x_{ij} & \geq 0 \ \ \ \ \ \ \mbox{for all jobs $j$ and machines $i$, $i \in S_j$}
\end{align*}

The dual LP is as follows (variables are $\alpha_j$ and $\beta_{(I,i,w)}$):
\begin{align*}\max \ \ \ \ \ \ \ \ &  \sum_j \alpha_j  - \sum_{(I,i,w)} \length(I) \cdot \beta_{(I,i,w)}\\\frac{\alpha_j}{p_j} & \leq \sum_{(I,i,w): r_j \in I, 2^w\leq w_j} \beta_{(I,i,w)} \ \ \ \ \
\mbox{for all jobs $j$ and machines $i\in S_j$} \\
\sum_{(I,i,w)} \frac{\beta_{(I,i,w)}}{2^w} & \leq 1 \\
\alpha_j, \beta_{(I,i,w)} & \geq 0
\end{align*}
The lemma now follows by setting $\beta_{(I,i,w)}$ to 1 if  $(I,i,w) \in \bI$, 0 otherwise.
\end{proof}

Now we bound the weight of jobs rejected by $\A$. We fix a time $t^\star$ -- this is the time by which all the jobs finish processing.
For rest of the discussion, we shall also fix a density class $d^\star$. We shall first bound the total weight of jobs of density
class $d^\star$ which are rejected by the algorithm $\A$. Finally, we shall take a sum over all values of $d^\star$ to bound the
total weight of jobs rejected by $\A$.

We begin with some definitions.
 Define $\Delta=\log(\alpha)$ (assume wlog that $\alpha$ is a power of 2). We shall be interested in jobs whose density class
 lies in the range $[\ds, \ds+\Delta]$.
For an integer $l$, let $\gamma_{l}$ denote $4l\alpha \opt.$ We will ensure that $\alpha/8 \leq l \leq \alpha/4$, and hence,
$\gamma_{l}$ will lie in the range $[ \alpha^2  \opt/2, \alpha^2 \opt]$. For a machine $i$, density class $d$ and time $t$,
let $\load_{i,d}(t)$ denote the maximum, over all pairs $(w,d)$  of
the total  weighted remaining processing time of jobs  in  the queue $Q_{i,w,d}$, i.e.,
$$\load_{i,d}(t) = \max_{w} \load_{i,w,d}(t). $$

Consider a job $j$ of type $(w,d)$ which gets dispatched to machine $i$. We define $\level(j)$ as $\load_{i,w,d}(r_j)+ w_j\cdot p_j$ --
this is the load it sees on machine $i$ (including its own weighted processing size).
For a  parameter $l$, we define a set of jobs  $\jobg{l}{\ds}$ as follows: a job $j$ of density class $d \in [\ds,\ds+\Delta]$ lies belongs to
$\jobg{l}{\ds}$ iff $\level(j) \geq \frac{\gamma_{l}}{2^{d-\ds}}$.

For each machine $i$ and parameter $l$ lying the range as
mentioned above, we define a set of disjoint weighted machine intervals $\intg{i}{l}{\ds}$.
 The procedure for this is given in Figure~\ref{figf:intwg}.  For a machine $i$ and time $t$,
$\maxload_{i,\ds}(t)$ is defined as $\max_{d \in [\ds,\ds+\Delta]}  2^{d-\ds} \cdot \load_{i,d}(t)$.

\begin{figure}[ht]
   \begin{center}
     \begin{boxedminipage}{6.5in}
          {\bf Constructing $\intg{i}{l}{\ds}$:} \medskip\\
        \sp \sp 1.   Initialize $t$ as $t^\star$ and  $\intg{i}{l}{\ds} \leftarrow \emptyset$ \\
         \sp \sp 2. {\bf Repeat} \\
         \sp \sp \sp \sp (i) Let $t_2$ be the highest time before (or equal to) $t$ such that $\maxload_{i,\ds}(t_2) \geq  \gamma_{l}$. \\
         \sp \sp \sp \sp (ii) Let $t_1$ be the earliest time such that $\maxload_{i,\ds}(t') >   \gamma_{l-1} $ for all
                              $t' \in (t_1, t_2]$. \\
         \sp \sp \sp \sp (iii) Let $w$ be the smallest weight class of a job $j \in \jobg{l-1}{\ds}$ such that $r_j \in (t_1, t_2]$ and $i \in S_j$. \\
         \sp \sp \sp \sp (iv) Add the weighted machine interval $([t_1, t_2],i,w)$ to $\intg{i}{l}{\ds}$. \\
         \sp \sp \sp \sp (v) $t \leftarrow t_1$. \\
         \sp \sp \sp \sp {\bf Until} No more intervals can be added to $\intg{i}{l}{\ds}$.
      \end{boxedminipage}
       \caption{Construction of the set of intervals in $\intg{i}{l}{\ds}$ for a machine $i$ and parameter $l$.}
       \label{figf:intwg}
       \end{center}
 \end{figure}


We now prove the analogue of Lemma~\ref{lemf:prop}.
\begin{lemma}
\label{lemf:propwg}
The sets of intervals in $\intg{i}{l}{\ds}$ satisfy the following properties:
\begin{itemize}
\item {\em (covering)}Consider a  job $j \in \jobg{l}{\ds}$ and a machine $i \in S_j$
Then there exists a weighted machine interval $(I,i,w) \in \intg{i}{l-1}{\ds}$
such that $r_j \in I$ and $w_j\geq 2^{w}$.
\item {\em (nesting)} Given any weighted machine interval $(I_l,i,w) \in \intg{i}{l}{\ds}$, there exists a weighted machine interval $(I_{l-1},i,w') \in \intg{i}{l-1}{\ds}$ for some $w' \leq w$
such that $I_{l-1}$ contains $I_l$.
\item {\em (processing)} For any weighted machine interval $(I,i,w) \in \intg{i}{l}{\ds}$,
the total duration in $I$ during which  jobs of density class in  $[\ds, \ds+\Delta]$ are processed on machine $i$ is at most the
total processing size of jobs in $\jobg{l-1}{\ds} \cap \{j: w_j \geq 2^{w}, r_j \in I\}$ which get dispatched to $i$ during $I$.
\end{itemize}
\end{lemma}
\begin{proof}
 Consider a job $j \in \jobg{l}{\ds}$ of  type $(w,d)$,  and a machine $i \in S_j$:
$$\load_{i,d}(r_j) \geq \load_{i,w,d}(r_j) \geq \level(j) - w_j\cdot p_j \geq \frac{\gamma_{l}}{2^{d-\ds}} - 4\opt \geq  \frac{\gamma_{l}}{2^{d-\ds}} - \frac{4\alpha\opt}{2^{d-\ds}} = \frac{\gamma_{l-1}}{2^{d-\ds}},$$ where the second inequality follows by our dispatch rule and the third inequality follows from the fact that $\opt\geq w_j\cdot p_j$ for any $j$. Therefore, $\maxload_{i,\ds}(r_j) \geq \gamma_{l-1}$.
Suppose, for the sake of contradiction that there
is no weighted machine interval in $\intg{i}{l-1}{\ds}$ containing $r_j$. Let $I_f$ be the first interval (if any) in $\intg{i}{l-1}{\ds}$
to the right of $r_j$, and suppose $s$ is the starting time of $I_f$ (if $I_f$ does not exist, $s$ is defined
as $t^\star$). After adding $I_f$ to $\intg{i}{l-1}{\ds}$, our algorithm
for constructing $\intg{i}{l-1}{d}$ would have tried
$t_2=r_j$, and hence there should be a weighted machine interval in  $\intg{i}{l-1}{\ds}$ containing $r_j$. This is a contradiction.
Therefore, there is a $(I,i,w') \in \intg{i}{l-1}{\ds}$ containing $r_j$. Since $j \in \jobg{l-1}{\ds}$, and $r_j \in I, i \in S_j$,
it must be the case that $w' \leq w$. This proves the first part of the lemma.

The argument above shows that if for any time $t$, machine $i$ and density class $d \in [\ds, \ds+\Delta]$,
$\load_{i,d}(t) \geq  \frac{\gamma_{l}}{2^{d-\ds}}$,
then there exists a weighted machine interval
$(I,i,w) \in \intg{i}{l-1}{\ds}$ containing $t$.  Suppose $(I_l=(s_l,e_l),i,w) \in \intg{i}{l}{\ds}$.
Since $\maxload_{i,\ds}(e_l) \geq \gamma_{l}$, there is a density class $d \in [\ds, \ds+\Delta]$ such that
$\load_{i,d}(e_l) \geq  \frac{\gamma_{l}}{2^{d-\ds}}$. Hence, there exists
 $(I_{l-1}=(s_{l-1},e_{l-1}),i, w') \in \intg{i}{l-1}{\ds}$ containing $e_l$.
 Since $\maxload_{i,\ds}(t) > \gamma_{l-1} $ for all $t \in (s_l, e_l)$, the left end-point of
$I_{l-1}$ will appear before $s_l$. Moreover, $w' \leq w$ (follows from the definition of $w$ or $w'$ in Step 2(iii) of
Figure~\ref{figf:intwg} and the fact that $I_{l}$ is contained in $I_{l-1}$). This proves the second part of the lemma.

It remains to prove the third part. Consider a weighted machine interval $(I, i,w) \in \intg{i}{l}{\ds}$.
\begin{claim}
\label{clmf:wqprocess}
Suppose we process a job $j$ of  type $(w,d)$ on machine $i$ at a particular time $t \in I$. Then it must be the case
that $d \geq \ds$ and  $$\load_{i,d}(t)=\load_{i,w,d}(t) > \frac{\gamma_{l-1}}{2^{d-\ds}}. $$
\end{claim}
\begin{proof}
Suppose $d < \ds$. We know that there is a density class $d' \in [\ds, \ds+\Delta]$ such that $\load_{i,d'}(t) > \frac{\gamma_{l-1}}{2^{d'-\ds}}$. The job processing rule for
$\A$ implies that $2^d \cdot \load_{i,w,d}(t) \geq 2^{d'} \cdot \load_{i,d'}(t) > 2^{\ds}  \gamma_{l-1}, $ and so, $ \load_{i,w,d}(t) > 2 \gamma_{l-1} \geq \alpha^2 \opt$, because
$\gamma_{l-1} \geq \alpha \opt/2$. But this is a contradiction (because of the job dispatch policy of $\A$). Hence, $d \geq \ds$.

If $\load_{i,d}(t) > \load_{i,w,d}(t)$, then let $w'$ be the weight class for which $\load_{i,d}(t) = \load_{i,w',d}(t)$. But then $2^d \cdot \load_{i,w,d}(t)
< 2^d \cdot \load_{i,w',d}(t)$, and so $\A$ cannot process $j$ on machine $i$ at time $t$. Similarly, if $d'$ is as above, then it must be the case that
$$ 2^d \cdot \load_{i,w,d}(t) \geq 2^{d'} \cdot \load_{i,d'}(t) > \gamma_{l-1} \cdot 2^{\ds}. $$ This implies the claim.
\end{proof}

Now consider a density class $d \in [\ds, \ds+\Delta],$ and a weight class $w'$.
  Let $t_f$ be the first  time in $I$ at which $\A$ processes a job $j$ of type $(w',d)$. The claim above shows that
$\load_{i,d}(t_f)=\load_{i,w',d}(t_f) > \frac{\gamma_{l-1}}{2^{d-\ds}}.$ 
 Further, $\load_{i,w',d}(t)$ remains at least $\frac{\gamma_{l-1}}{2^{d-\ds}}$
after $t_f$ till the end of the interval $I$ -- indeed, Claim~\ref{clmf:wqprocess} says that if  $\load_{i,w',d}(t)$ is at most $\frac{\gamma_{l-1}}{2^{d-\ds}}$ for some time $t \in I$,
then the algorithm does not process a job of type $(w',d)$ at this time on machine $i$.
 Hence, $\load_{i,w',d}(t)$ will not go below $\frac{\gamma_{l-1}}{2^{d-\ds}}$ after $t_f$ (during $I$).
Also, for any time $t$ during $[t_1,t_f)$, $\load_{i,w',d}(t) < \frac{\gamma_{l-1}}{2^{d-\ds}}$, where $t_1$ is the left end-point of $I$ (by definition of $t_f$ and the fact that this statement holds for $t_1$).
 Let $V$ denote the total volume of time during $I$ when we process a job of type $(w',d)$. Then it must happen that at least $V$ volume of jobs 
 of type $(w',d)$ are released during $[t_f, t_e]$, where $t_e$ is the right end-point of $I$ -- if this does not
happen then we will end up processing a job of type $(w',d)$ at a time $t$ in $I$ even when $\load_{i,w',d}(t) < \frac{\gamma_{l-1}}{2^{d-\ds}}$, a contradiction. For any job $j'$ of
type $(w',d)$ released during $[t_f, t_e]$, $\load_{i,w',d}(r_j) \geq \frac{\gamma_{l-1}}{2^{d-\ds}}$, and so, $j' \in \jobg{l-1}{\ds}$. Recall that the weighted machine interval
was denoted by $(I,i,w)$ -- by definition of $w$, it must be satisfy $w \leq w'$. Summing over all $w'$ gives us the lemma. 
\end{proof}

We now define a set of weighted machine intervals $\bI$ and dual values $\alpha_j$ for all jobs $j$ which will satisfy the conditions of Lemma~\ref{lemf:genlower}. Let $\bI$ be the
set of weighted machine intervals defined above, i.e., $\cup_{l=\alpha/8}^{\alpha/4} \cup_i \intg{i}{l}{\ds}$ -- note that this
is a multi-set, i.e., if a weighted machine interval $(I,i,w)$ appears in several of the sets $\intg{i}{l}{\ds}$, it is counted these
many times.
For each job $j \in \cup_{l=\alpha/8}^{\alpha/4} \jobg{l}{\ds}$ which does not get rejected,
 define $\alpha_j = p_j \cdot (l_j-\alpha/8)$, where $l_j$ is the largest value $l$ such that $j  \in
 \jobg{l}{\ds}$. Further, if $j$ is a job of density class $\ds$  gets rejected,  we set $\alpha_j = p_j \cdot \alpha/8$. For remaining
 jobs, we set $\alpha_j=0$.


 \begin{claim}
 \label{clf:condition-genwt}
 The set of intervals $\bI$ and the values $\alpha_j$ defined above satisfy the feasibility conditions~\eqref{eqf:dualcond}.
 \end{claim}

\begin{proof}
Consider a job $j$ for which $\alpha_j = p_j \cdot (l_j-\alpha/8), $ and a machine $i \in S_j$. Lemma~\ref{lemf:propwg} shows that
there are weighted machine intervals $(I_l,i,w_l) \in \intg{i}{l}{\ds}$ for $l=\alpha/8, \ldots, l_j-1$, such that $r_j \in I_l$
and $2^{w_l} \leq w_j$. Thus, the conditions~\eqref{eqf:dualcond} are satisfied for $j$.
Similarly, if $j$ is a job of density class $\ds$ which gets rejected, and $i \in S_j$, then $\level(j) \geq \alpha^2 \opt
\geq \gamma_l$, for $l=\alpha/4$. Thus, $j \in \jobg{\alpha/4}{\ds}$, and so, the same argument as above applies
here as well.
\end{proof}

Claim~\ref{clf:condition-genwt} implies that we can apply Lemma~\ref{lemf:genlower}. We give some notation first.
For a parameter $l$ and weight class $w$, let $\bint{l}{w}{\ds}$ denote the following set of weighted machine intervals:
$$\{ (I,i,w):  (I,i,w) \in \cup_{i'} \intg{i'}{l}{\ds}\}. $$

Let $m_{l,w,\ds}$ denote $|\bint{l}{w}{\ds}|$, and $\len{l,w,\ds}$ denote the total length of  (associated intervals in) the weighted machine intervals in
$\bint{l}{w}{\ds}$.
Applying Lemma~\ref{lemf:genlower}, we get
\begin{eqnarray}
\label{eqf:lengthBound}
\opt \cdot \sum_{w}\sum_{l=\alpha/8}^{\alpha/4} \frac{m_{l,w,\ds}}{2^w} \geq \sum_j \alpha_j - \sum_{w}\sum_{l=\alpha/8}^{\alpha/4} \len{l,w,\ds}.
\end{eqnarray}

\begin{claim}
\label{clmf:lengthwg}
Let $\jrej(\ds)$ denote the jobs of density class $\ds$ which get rejected. Also let $P_{\geq \ds+\Delta}$ denote the total processing time of jobs of density class higher than $\ds+\Delta$. Then,
$$ \sum_{j \notin \jrej(\ds)} \alpha_j \geq \sum_{w}\sum_{l=\alpha/8+1}^{\alpha/4} \len{l,w,\ds} - \alpha/8 \cdot P_{\geq \ds+\Delta}. $$
\end{claim}
\begin{proof}
We split $\len{l,w,\ds}$ into two parts -- let $\len{l,w,\ds}'$ denote the volume during intervals in $\bint{l}{w}{\ds}$ where $\A$ processes a job of 
density class lying in the set $[\ds, \ds+\Delta]$, and
$\len{l,w,\ds}''$ be the volume during intervals in $\bint{l}{w}{\ds}$ where $\A$ processes a job of class density higher than $\ds + \Delta$. Claim~\ref{clmf:wqprocess} implies that
$\len{l,w,\ds}=\len{l,w,\ds}'+\len{l,w,\ds}''$.

Lemma~\ref{lemf:propwg}~(third part) implies that $\sum_w \len{l,w,\ds}' \leq \sum_{j \in \jobg{l-1}{\ds}, j \notin \jrej(\ds)} p_j$ (the intervals in $\bint{w}{l}{\ds} \cap \intg{i}{l}{\ds}$ are disjoint).
Summing over $l=\alpha/8+1, \ldots, \alpha/4$, we get
$$ \sum_{l=\alpha/8+1}^{\alpha/4} \sum_w \len{l,w,\ds}' \leq \sum_{l=\alpha/8+1}^{\alpha/4}\sum_{j \in \jobg{l-1}{\ds},j \notin \jrej(\ds) } p_j =\sum_{j \notin \jrej(\ds)} p_j \cdot(l_j-\alpha/8) = \sum_{j \notin \jrej(\ds)} \alpha_j $$

Now we bound $\len{l,w,\ds}''$. For a fixed $l$, $\sum_{w} \len{l,w,\ds}''$ is at most $P_{\geq \ds+\Delta}.$ Thus,
$$\sum_{l=\alpha/8+1}^{\alpha/4} \sum_w \len{l,w,\ds}'' \leq \alpha/8 \cdot P_{\geq \ds+\Delta}.$$
Combining the above inequalities gives us the desired result.
\end{proof}

Hence, inequality~\eqref{eqf:lengthBound} can be simplified as
\begin{eqnarray}
\label{eqf:lowerB2}
\opt \cdot \sum_{w}\sum_{l=\alpha/8}^{\alpha/4} \frac{m_{l,w, \ds}}{2^w} + \alpha/8 \cdot P_{\geq \ds+\Delta} + \sum_w \len{\alpha/8,w,\ds} \geq \sum_{j \in \jrej(\ds)} \alpha_j  = \sum_{j \in \jrej(\ds)}\frac{\alpha \cdot p_j}{8}
\end{eqnarray}
We simplify the above expression to bound the weight of jobs of density class $\ds$ which get rejected.
Some more notation. For a weight class $w$ and density class $d$, let $V_{w,d}$ denote the total weighted processing size of jobs of type $(w,d)$,
and let $W_{w,d}$ be the total weight of such jobs, i.e.,
$$ V_{w,d} = \sum_{j: \type(j)=(w,d)} w_j p_j, \ \ W_{w,d} = \sum_{j: \type(j)=(w,d)} w_j. $$

For parameters $a,b, a \leq b$, let $V_{w,d}(a,b)$ and $W_{w,d}(a,b)$ denote the corresponding sum for jobs $j$ with $\level(j) \in [a,b]$, i.e,
$$ V_{w,d}(a,b) = \sum_{j:\type(j)=(w,d), \level(j) \in [a,b]} w_j p_j, \ \ W_{w,d}(a,b) = \sum_{j:\type(j)=(w,d), \level(j) \in [a,b]} w_j. $$

\begin{claim}
\label{clf:countwg}
\begin{align*}
\opt \cdot \sum_{w}\sum_{l=\alpha/8}^{\alpha/4} \frac{m_{l,w,\ds}}{2^w} + \sum_w \len{\alpha/8,w, \ds}
 \leq \sum_{d \in [\ds, \ds+\Delta]} \sum_w \frac{1}{2 \alpha 2^{\ds}}  W_{w,d}  +  P_{\geq \ds},
\end{align*}
where $P_{\geq \ds}$ denote the total processing time of jobs of density class at least $\ds$. 
\end{claim}

\begin{proof}
First we consider $\sum_w \len{\alpha/8,w,\ds}$ -- this is just the total length of the weighted machine intervals $\cup_i \intg{i}{\alpha/8}{\ds}$. Note that these weighted machine intervals are disjoint and  Lemma~\ref{lemf:propwg} implies that we will only process jobs of
density class at least $\ds$ during these intervals.   Hence,
$\sum_ \len{\alpha/8,w,\ds}$ is at most $P_{\geq \ds}$.

Finally, we consider the term $\opt \cdot \sum_{w}\sum_{l=\alpha/8}^{\alpha/4} \frac{m_{l,w,\ds}}{2^w}$. We shall upper bound each term of the summation as follows.
 Consider a machine interval $(I,i,w) \in \intg{i}{l}{\ds}$. Let $I=[s,t]$. We know that there exits a pair $(w,d)$ 
  such that $\load_{i,w,d}(s) < \frac{\gamma_{l-1}}{\powd{d}},$ but $\load_{i,w,d}(t) \geq \frac{\gamma_{l}}{\powd{d}}$. Let $j_1$ be
 the first job of type $(w,d)$ dispatched to $i$ during $I$ for which $\level(j_1) \geq \frac{\gamma_{l-1}}{\powd{d}}$ and $j_2$ be the first
 such job for which $\level(j_2) \geq \frac{\gamma_{l}}{\powd{d}}$. All jobs of type $(w,d)$ dispatched to $i$ after $j_1$
 and before $j_2$ will have level in the range $\left(\frac{\gamma_{l-1}}{\powd{d}},\frac{\gamma_{l}}{\powd{d}} \right)$, because as
 argued in the proof of Lemma~\ref{lemf:propwg}(third part), once $\load_{i,w,d}(t)$ goes above $\frac{\gamma_{l}}{\powd{d}}$, it will never go below
 this quantity till the end of $I$. Further
 the total weighted volume of such jobs will be at least $$\frac{\gamma_{l}}{\powd{d}}-\frac{\gamma_{l-1 }}{\powd{d}}-w_{j_1}p_{j_1} - w_{j_2}p_{j_2}
 \geq \frac{4\alpha \opt}{\powd{d}} - 2\opt \geq \frac{2\alpha \opt}{\powd{d}} ,$$
 where the last inequality follows from the fact that $\alpha \geq \powd{d}$. Thus, we get
 $$  \frac{\opt}{2^w} \leq \frac{\powd{d}}{ 2\alpha 2^w} \cdot \sum_{j: r_j \in I, \type(j)=(w,d), \level(j) \in \left(\frac{\gamma_{l-1}}{\powd{d}},\frac{\gamma_{l}}{\powd{d}} \right)} w_j p_j$$
 Summing over all weighted machine intervals in $\cup_i \intg{i}{l}{\ds}$, we get
 $$ \sum_w \frac{\opt \cdot m_{l,w,\ds}}{2^w} \leq  \sum_{d \in [\ds, \ds+\Delta]} \sum_{w} \frac{\powd{d}}{2 \alpha 2^w} \cdot V_{w,d}  \left(\frac{\gamma_{l-1}}{\powd{d}},\frac{\gamma_{l}}{\powd{d}} \right).$$
 Summing over all $l$ and using the fact that if $a \leq b \leq c$, then $V(a,b) + V(b,c)=V(a,c)$, we get
\begin{eqnarray*}
 \sum_{l=\alpha/8}^{\alpha/4} \sum_w \frac{\opt \cdot m_{l,w,\ds}}{2^w} & \leq & \sum_{d \in [\ds, \ds+\Delta]} \sum_{w} \frac{\powd{d}}{2 \alpha 2^w}
\cdot V_{w,d}  \left(\frac{\gamma_{\alpha/8-1}}{\powd{d}},\frac{\gamma_{\alpha/4}}{\powd{d}} \right) \\
& \leq & \sum_{d \in [\ds, \ds+\Delta]} \sum_{ w} \frac{\powd{d}}{2 \alpha 2^w}
\cdot V_{w,d} \ \leq \  \sum_{d \in [\ds, \ds+\Delta]} \frac{1}{2\alpha 2^{\ds}} \cdot \sum_{w} W_{w,d},
\end{eqnarray*}
where the last inequality follows from the fact that for any job $j$ of type $(w,d)$ $\frac{2^w}{2^{d+1}} < p_j\leq \frac{2^w}{2^{d}}$ and so, 
$$ \frac{2^d}{2^w} V_{w,d} \leq  \sum_{j: \type(j)=(w,d)} w_j/p_j \cdot p_j = W_{w,d}.$$
 Thus we have shown the desired result.
 \end{proof}

We are now ready to bound the weight of jobs rejected by $\A$.
\begin{lemma}
\label{lemf:finalA}
The total weight of jobs rejected by $\A$ is at most $\eps$ times the total weight of all the jobs, provided we pick $\alpha=\frac{76}{\eps}$.
\end{lemma}
\begin{proof}
We first bound the total weight of jobs of density class $\ds$ which get rejected by $\A$.  We have
\begin{align*}
\sum_{j \in \jrej(\ds)} w_j & \leq  2^{\ds+1} \cdot \sum_{j \in \jrej(\ds)} p_j \ \stackrel{(\ref{eqf:lowerB2})}{\leq} \
\frac{16 \cdot 2^{\ds} \opt}{\alpha}  \cdot \sum_{w}\sum_{l=\alpha/8}^{\alpha/4} \frac{m_{l,w, \ds}}{2^w} + 2^{\ds+1} \cdot P_{\geq \ds+\Delta} + \frac{16 \cdot 2^{\ds}}{\alpha} \sum_w \len{\alpha/8,w,\ds} \\
& \stackrel{Claim~\ref{clf:countwg}}{\leq}  \sum_{d \in [\ds, \ds+\Delta]} \sum_{w} \frac{8}{ \alpha^2}  W_{w,d}  +  \frac{16 \cdot 2^{\ds}}{\alpha}P_{\geq \ds}
+  2^{\ds+1} \cdot P_{\geq \ds+\Delta} \\
& = \frac{8}{ \alpha^2}  \cdot \sum_{d=\ds}^{\ds+\Delta} \sum_{j \in J_d} w_j + \frac{16 \cdot 2^{\ds}}{\alpha} \sum_{d \geq \ds} \sum_{j \in J_d} p_j + 2^{\ds+1} \sum_{d \geq \ds+\Delta}
\sum_{j \in J_d} p_j
\end{align*}
where $J_d$ denotes the jobs of density class $d$. Summing over all values of $\ds$, the total weight of rejected jobs can be expressed as
\begin{align*}
\sum_{\ds} \sum_{j \in \jrej(\ds)} w_j & \leq \frac{8}{\alpha^2} \sum_{\ds} \sum_{d=\ds}^{\ds+\Delta} \sum_{j \in J_d} w_j + \sum_{\ds} \frac{16 \cdot 2^{\ds}}{\alpha} \sum_{d \geq \ds} \sum_{j \in J_d} p_j + \sum_{\ds} 2^{\ds+1} \sum_{d \geq \ds+\Delta}
\sum_{j \in J_d} p_j \\
& \leq  \frac{8 \Delta}{\alpha^2} \sum_d \sum_{j \in J_d} w_j + \sum_d \sum_{j \in J_d} p_j \sum_{\ds \leq d}  \frac{16 \cdot 2^{\ds}}{\alpha} + \sum_d \sum_{j \in J_d} p_j \sum_{\ds \leq d-
\Delta} 2^{\ds+1} \\
& \leq   \frac{8 \Delta}{\alpha^2} \sum_j w_j + \frac{64}{\alpha} \sum_d \sum_{j \in J_d} p_j 2^{d+1} + \frac{4}{\alpha} \sum_d \sum_{j \in J_d} p_j 2^d \\
& \leq \left(  \frac{8 \Delta}{\alpha^2}  + \frac{64}{\alpha}  + \frac{4}{\alpha} \right) \sum_j w_j \ \leq \ \eps \cdot \sum_j w_j,
\end{align*}
if we pick $\alpha=\frac{76}{\eps}$.
\end{proof}

Thus we have shown the main theorem of this section.
\begin{theorem}
\label{thmf:A}
The algorithm $\A$ rejects jobs of total weight at most $\eps$ times the total weight of all jobs, and ensures that for any machine $i$, time $t$, and pair $(w,d)$, the total weighted remaining processing time of jobs of type $(w,d)$ at time $t$ on machine $i$ is at most $\alpha^2 \opt$. Further, $\A$ is an immediate dispatch algorithm which rejects jobs on arrival only.
\end{theorem}

\noindent
{\bf Analysis for algorithm $\B$:} Now we analyze the algorithm $\B$. It is clear that the weighted flow-time of any job $j$ is at most $2\beta \opt,$ where $\beta$ denotes $\frac{\alpha^2}{\eps^2}$. We need to bound the weight of jobs
rejected by $\B$. We can restrict our attention to a fixed machine because $\B$ processes the same set of jobs on a machine as $\A$ does. Further, $\A$ does not reject any job once it
gets dispatched to a machine. For rest of the discussion we fix a machine $\is$, and bound the weight of jobs which were dispatched by $\A$
to $\is$, but got rejected by $\B$. We also fix a density class $\ds$ and first bound the jobs of density class $\ds$ which get rejected
by $\B$. Let $\jobB{\is}{\ds}$ denote the set of jobs of density class $\ds$ which are dispatched by $\A$ to the machine $\is$.

\begin{figure}[ht]
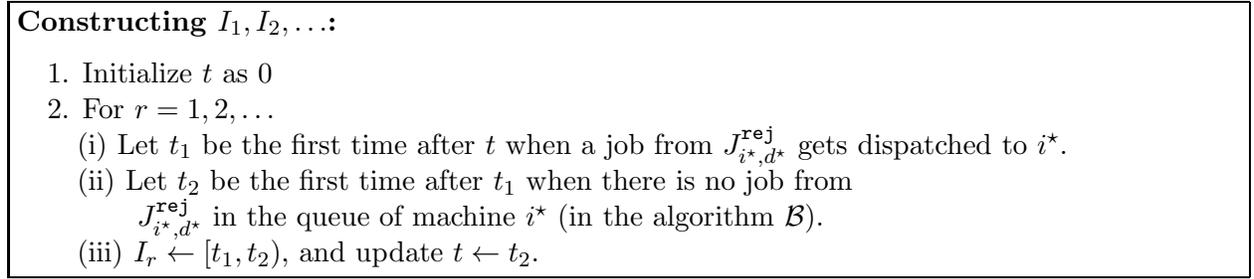

   \begin{center}
     \begin{boxedminipage}{6.5in}
          {\bf Constructing $I_1, I_2, \ldots$:} \medskip\\
        \sp \sp 1.   Initialize $t$ as 0 \\
         \sp \sp 2. For $r=1,2,\ldots$ \\
         \sp \sp \sp \sp (i) Let $t_1$ be the first time after $t$ when a job from $\jobrejB{\is}{\ds}$ gets dispatched to $\is$. \\
         \sp \sp \sp \sp (ii) Let $t_2$ be the first time after $t_1$ when there is no job from \\
         \sp \sp \sp \sp \sp \sp \sp \sp $\jobrejB{\is}{\ds}$ in the queue of machine $\is$ (in the algorithm $\B$). \\
         \sp \sp \sp \sp (iii) $I_r \leftarrow [t_1, t_2)$, and update $t \leftarrow t_2$.
      \end{boxedminipage}
       \caption{Construction of the set of intervals $I_1, I_2, \ldots$.}
       \label{figf:formintervals}
       \end{center}
 \end{figure}

Let $\jobrejB{\is}{\ds}$ denote the subset of jobs in $\jobB{\is}{\ds}$ which get rejected by $\B$. First we divide the time line into
 disjoint intervals $I_1, I_2, \ldots$ with the following properties: (i) for any $j \in \jobrejB{\is}{\ds}$, there is an interval
$I_r$ which contains the time period when $j$ was waiting in the queue of machine $\is$ (i.e., the time period from $r_j$ to the time when
it gets rejected), (ii) for any interval $I_r$ and time $t \in I_r$, there is a job from $\jobrejB{\is}{\ds}$ which is waiting in the
queue of machine $\is$. We can easily form these intervals by a greedy procedure. For sake of completeness, this procedure is described in
Figure~\ref{figf:formintervals}.

For an interval $I_r$, let $\jobrejB{\is}{\ds}(I_r)$ denote the jobs in $\jobrejB{\is}{\ds}$ which are released during $I_r$.
Define $\jobB{\is}{\ds}(I_r)$ similarly.  We fix an
interval $I_r$ and bound the weight of jobs in $\jobrejB{\is}{\ds}(I_r)$. Let $w_{\min}(I_r)$ denote the smallest weight class of a job in
$\jobrejB{\is}{\ds}(I_r)$. For a time $t$ and density class $d$, we define an indicator variable $\bone{\A}{i}{d}(t)$ which is 1 iff
$\A$ processes a job of density class $d$ at time $t$ on machine $i$. Define $\bone{\B}{i}{d}(t)$ similarly.
Let $\proc{\A}{\is}{\ds}(I_r)$
be the total volume of processing of density class $\ds$ jobs performed by $\A$ during $I_r$ on machine $\is$, i.e.,
$$\proc{\A}{\is}{\ds}(I_r) = \int_{t \in I_r}  \bone{\A}{\is}{\ds}(t) dt. $$

\begin{claim}
\label{clf:Aproc}
The total processing size of jobs in $\jobB{\is}{\ds}(I_r)$ whose weight class is at least $w_{\min}(I_r)$ is at most
$\proc{\A}{\is}{\ds}(I_r)+ \frac{2 \alpha^2 \opt}{2^{w_{\min}(I_r)}}.$
\end{claim}
\begin{proof}
In the schedule $\A$, the jobs in $\jobB{\is}{\ds}(I_r)$ will either get processed during $I_r$ or will appear in the queue of machine $\is$ at the end of $I_r$. The total volume of the
former quantity is at most $\proc{\A}{\is}{\ds}(I_r)$. For the latter quantity, observe that we are interested in weight classes $w_{\min}(I_r)$ and higher. Theorem~\ref{thmf:A}
shows that this quantity can be at most
$$ \sum_{w \geq w_{\min}(I_r)} \frac{\alpha^2 \opt}{2^{w}} \leq \frac{2 \alpha^2 \opt}{2^{w_{\min}(I_r)}}.$$
\end{proof}

Define another indicator variable $\bone{}{\is}{\ds}(t)$ which is 1 iff both $\bone{\A}{\is}{\ds}(t)$ and $\bone{\B}{\is}{\ds}(t)$ are 1 (the absence
of superscript means it is applied to both $\A$ and $\B$). Again, define $\proc{}{\is}{\ds}(I_r)$ as the volume of time during $I_r$ for which
$\bone{}{\is}{\ds}(t)=1$. Note a few important points: (i) If $\bone{}{\is}{\ds}(t)=1$, for some $t \in I_r$,
 then $\B$ processes a job of density class  $\ds$ and
weight class at least $w_{\min}(I_r)$ at time $t$, because among all jobs of density $\ds$, it prefers jobs of higher weight, and there is always a
job of weight class at least least $w_{\min}(I_r)$ waiting in the queue of machine $\is$, (ii) If $\bone{\A}{\is}{\ds}(t)=1, t \in I_r$ and $\bone{\B}{\is}{\ds}(t)=0,$
then $\B$ processes a job of density class  at least  $\ds + \logeps$ at time $t$.

Now, we want to disregard the part of $\proc{}{\is}{\ds}(I_r)$ where $\B$ processes a job which eventually gets rejected. We say that a time
$t$ satisfying $\bone{}{\is}{\ds}(t)=1$ is {\em bad} if $\B$ processes a job from $\jobrejB{\is}{\ds}(I_r)$ at time $t$ on $\is$. We first
show that the total volume of bad time is small.

\begin{lemma}
\label{lemf:waste}
The total volume of bad time in $I_r$ is at most $\frac{8 \length(I_r)}{\beta}$.
\end{lemma}
\begin{proof}
For each weight class $w \geq w_{\min}(I_r)$, we bound the volume of bad time at which $\B$ is processing a job of weight class $w$.
Note that for a job of density class $\ds$ and weight class $w$, its processing time lies in the range $[2^{w-\ds}, 2^{w-\ds-1})$.
Let $p_w$ denote $w-\ds$.

Fix a weight class $w$. Recall that the algorithm $\B$ divides the time line into segments of length $\beta \opt/2^w$. Consider such a segment
$S$ which intersects with $I_r$. Let $S_L$ be the segment to the left of $S$. Any job of weight class $w$ processed by
$\B$ during $S$ must have been released in $S$ or $S_L$ (if it were released earlier, $\B$ would have rejected it by the end of $S_L$).
Since $\B$ processes jobs in order of release dates, there will be at most one job $j$ for which $r_j \in S_L$ and $\B$ processes $j$ during
a bad time in $S$. If there were two such jobs $j$ and $j'$ (and say $r_j \leq r_{j'}$), then $\B$ would have completed $j$ before starting
$j'$ (note that $j$ can get rejected at the end of $S$ only). But then it could not have rejected $j$, a contradiction (recall that a job
processed during a bad time gets rejected). Hence, the total number volume of bad time in $I_r$ during which $\B$ processes a job of
weight class $w$ is at most the number of such segments which intersect $I_r$ times the maximum size of a job of weight class $w$ (and density
class $\ds$), i.e.,
$$   \left( \frac{\length(I_r)}{\beta \opt/2^w}+2 \right) \cdot 2^{p_w} \leq \frac{3 \cdot 2^{p_w} \cdot \length(I_r)}{\beta \opt/2^w} = \frac{3 \cdot 2^{2w} \length(I_r)}{\beta \opt  2^{\ds}},$$
where the first inequality follows from the fact that $I_r$ must contain at least one segment for the weight class $w_{\min}(I_r)$, and so,
$\length(I_r) \geq \frac{\beta \opt}{2^{w_{\min}(I_r)}} \geq \beta \opt/2^w$.

We sum over all $w \geq w_{\min}(I_r)$, and let $w_{\max}$ be the highest weight class among all jobs of density class $\ds$. The total
volume of bad time in $I_r$ can now be bounded as
$$ \sum_{w \leq w_{\max}} \frac{3 \cdot 2^{2w} \length(I_r)}{\beta \opt  2^{\ds}} \leq \frac{4 \cdot 2^{2w_{\max}} \length(I_r)}{\beta \opt  2^{\ds}} \leq \frac{8 \cdot \length(I_r)}{\beta},$$
where the last inequality follows from the fact that $\opt \geq 2^{w_{max}} \cdot 2^{w_{\max}-\ds-1}$ (the weighted size of a job of density
class $\ds$ and weight class $w_{\max}$).
\end{proof}

\begin{corollary}
\label{corf:countB}
The total processing time of density class $\ds$ jobs which are rejected by $\B$ during $I_r$ on machine $\is$ is at most
$$ \proc{\A}{\is}{\ds}(I_r)+  4 \cdot \eps^2 \length(I_r) - \proc{}{\is}{\ds}(I_r)$$
\end{corollary}

\begin{proof}
The result follows from combining Lemma~\ref{lemf:waste} and Claim~\ref{clf:Aproc}. Claim~\ref{clf:Aproc} shows that the total processing time of such
jobs is $\proc{\A}{\is}{\ds}(I_r)+ \frac{2 \alpha^2 \opt}{2^{w_{\min}(I_r)}}.$
minus the processing times of such jobs which complete processing in $\B$. Lemma~\ref{lemf:waste} shows that the latter quantity is at least
$\proc{}{\is}{\ds}(I_r)-\frac{8 \length(I_r)}{\beta}$. Therefore, the processing time of density class $\ds$ jobs which are rejected by $\B$ during $I_r$ on machine $\is$ is
at most
$$ \proc{\A}{\is}{\ds}(I_r)+ \frac{2 \alpha^2 \opt}{2^{w_{\min}(I_r)}} + \frac{8 \length(I_r)}{\beta} - \proc{}{\is}{\ds}(I_r).$$
Since $\length(I_r) \geq \frac{\beta \opt}{2^{w_{\min}(I_r)}} = \frac{\alpha^2 \opt}{\eps^2 2^{w_{\min}(I_r)}}  $, the result follows.
\end{proof}

\begin{theorem}
\label{thmf:B}
The total weight of jobs rejected by $\B$ is at most $20 \eps$ times the total weight of all jobs.
\end{theorem}
\begin{proof}
Fix an interval $I_r$. Corollary~\ref{corf:countB} gives the total weight of density $\ds$ jobs which get rejected. We write it in a form which will be more useful.
For a density class $d$ and time $t$, let $\bone{\B}{\is}{d}(t)$ be the indicator variable which is 1 iff $\B$ processes a job of density class $d$ at time $t$ on machine $\is$.
Similarly, define $\bone{\B}{\is}{\geq d}(t)$ to be 1 iff $\B$ processes a job of density class at least $d$ during $t$ on machine $\is$.

We first observe that
$$ \proc{\A}{\is}{\ds}(I_r)-\proc{}{\is}{\ds}(I_r) \leq \int_{t \in I_r}  \bone{\B}{\is}{\geq \ds+ \logeps}(t) dt,$$
because the LHS is 1 iff at time $t$, $\B$ processes a job of density class at least $\ds + \logeps$ at time $t$. Further,
$$\length(I_r) = \int_{t \in I_r} \bone{\B}{\is}{\geq \ds - \logeps}(t)  dt,$$
because if at any time $t \in I_r,$ there is a job of density class $\ds$ waiting in the queue at machine $i$ at time $t$, and so,
the processing rule for $\B$ dictates that it cannot
process a job of density class less than  $\ds - \logeps$ at time $t$.
 Combining the above two inequalities with Corollary~\ref{corf:countB}, the total weight of jobs of density $\ds$ rejected
during $I_r$ is at most
$$4 \eps^2 \cdot 2^{\ds+1} \int_{t \in I_r} \bone{\B}{\is}{\geq \ds - \logeps}(t) dt + 2^{\ds+1} \cdot \int_{t \in I_r}  \bone{\B}{\is}{\geq \ds + \logeps}(t) dt .$$
Summing the above for all intervals $I_r$ and noting that these intervals are disjoint for a fixed $\ds$, the total weight of jobs of density class $\ds$ rejected is at most 
$$4 \eps^2 \cdot 2^{\ds+1} \int_{t} \bone{\B}{\is}{\geq \ds - \logeps}(t) dt + 2^{\ds+1} \cdot \int_{t}  \bone{\B}{\is}{\geq \ds + \logeps}(t) dt .$$
Summing the above for all density classes $\ds$, we see that the total weight of jobs rejected by $\B$ is at most
\begin{align*}
& 4 \eps^2 \sum_{\ds} 2^{\ds +1} \int_{t} \sum_{d \geq  \ds - \logeps} \bone{\B}{\is}{d}(t) dt + \sum_{\ds} 2^{\ds+1} \cdot \int_{t}  \sum_{d \geq \ds+ \logeps} \bone{\B}{\is}{d}(t) dt \\
& = 4 \eps^2 \int_{t} \sum_d \sum_{\ds \leq d + \logeps} 2^{\ds+1} \bone{\B}{\is}{d}(t) dt + \int_{t} \sum_{d} \sum_{\ds \leq  d - \logeps}
2^{\ds+1} \bone{\B}{\is}{d}(t) dt \\
& \leq 10 \eps \int_{t} \sum_d 2^d \bone{\B}{\is}{d}(t) dt
\end{align*}
Now note that $\sum_i \int_t \sum_d 2^d \bone{\B}{i}{d}(t) dt$ is at most twice the total weight of all jobs processed by $\B$ (a job of class density $d$
has density at most $2^{d+1}$). This proves the theorem.
\end{proof}
\subsection{Removing the assumption about $\opt$}
\label{secf:remlinf}
\begin{figure}[ht]
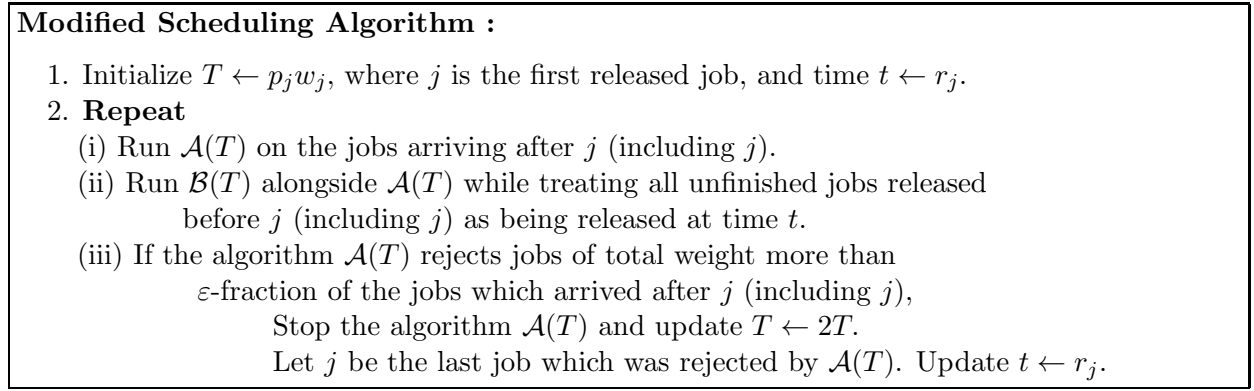

   \begin{center}
     \begin{boxedminipage}{6.5in}
          {\bf Modified Scheduling Algorithm :} \medskip\\
        \sp \sp 1.   Initialize $T \leftarrow p_j w_j,$ where $j$ is the first released job, and time $t \leftarrow r_j$. \\
         \sp \sp 2. {\bf Repeat} \\
         \sp \sp \sp \sp (i) Run $\A(T)$  on the jobs arriving after $j$ (including $j$). \\
         \sp \sp \sp \sp (ii) Run $\B(T)$ alongside $\A(T)$ while treating all unfinished jobs released \\
         \sp \sp \sp \sp \sp \sp \sp  \sp \sp \sp \sp  before $j$ (including $j$) as being released at time $t$. \\
         \sp \sp \sp \sp (iii) If the algorithm $\A(T)$ rejects jobs of total weight more than  \\
         \sp \sp \sp \sp \sp \sp \sp \sp \sp \sp \sp \sp $\eps$-fraction of the jobs which  arrived after $j$ (including $j$), \\
         \sp \sp \sp \sp \sp \sp \sp \sp \sp \sp  \sp \sp \sp \sp \sp \sp \sp Stop the algorithm $\A(T)$ and update $T \leftarrow 2 T. $ \\
          \sp \sp \sp \sp \sp \sp \sp \sp \sp \sp \sp  \sp \sp \sp \sp \sp \sp  Let $j$ be the last job which was rejected by $\A(T)$. Update $t \leftarrow r_j$.
      \end{boxedminipage}
       \caption{Scheduling algorithm without any assumption on $\opt$. }
       \label{figf:modA}
       \end{center}
 \end{figure}

So far we have assumed that we know the value of $\opt$. Now we explain how to get rid of this assumption. Our algorithm starts with an estimate for $\opt$
and updates it whenever we end up rejecting more than  the desired weight of jobs. For a parameter $T$, let $\A(T)$ denote the algorithm $\A$ when
the estimate for $\opt$ is given by $T$. Define $\B(T)$ similarly. The modification to the scheduling algorithm is described in Figure~\ref{figf:modA}. Note that when running $\A(T)$ in Step~2(i),
the algorithm $\A(T)$ completely ignores the jobs dispatched before $j$, and so, these jobs do not figure in the calculation of $\load$. In fact, these jobs
will never get processed by $\A(T)$. We shall refer to each iteration in Step~2 as a {\em phase}. When we run $\B(T)$ in Step~2(ii), we treat the
unfinished jobs of previous phase as being released at the beginning of this phase. So even though such jobs do not affect $\A(T)$ in the current phase,
$\B(T)$ may schedule them (or reject them).

We now analyze the scheduling algorithm.

\begin{theorem}
\label{thmf:final}
The above algorithm is $O(1/\eps^4)$-competitive, and rejects jobs of total weight $O(\eps)$ times the total weight  of  all jobs.
\end{theorem}
\begin{proof}
Let $\opt$ denote the value of the offline optimum.
Let $T_i$ be the value of $T$ at the beginning of phase $u$. First observe that in the last phase $u^\star$, $T_{u^\star} \leq 2 \opt$. Indeed, if $T_u$ becomes
larger than $\opt$, then $\A(T_u)$ will not reject jobs of weight more then $\eps$-fraction of the weight of all the jobs in this phase (the offline optimum
for jobs released in this phase can only be at most $\opt$).

Also, observe that $\A(T_u)$ rejects jobs of weight at most $\eps$-times the weight of all jobs released in this phase, and so, the total weight of jobs rejected
by it is within $\eps$-fraction of all the jobs. Now, observe that in  a phase $u$, the total load of jobs of type $(w,p)$ waiting in the queue of a machine $i$ at time $t$ is at most $$ \sum_{u' \leq u} \frac{\alpha^2 T_u}{2^w} \leq \frac{2 \alpha^2 T_u}{2^w}.$$ This follows from Theorem~\ref{thmf:A} about the
properties of $\A$. Thus, the previous phases worsen the queue size by a factor of 2 only.

Now, we consider $\B$. Suppose it completes a job $j$ in a phase $u$ which was released in phase $u' \leq u$. In a phase $u''$ between $u'$ and $u$,
$j$ could have waited for at most $\frac{2 \alpha^2 T_{u''}}{\eps^2 w_j}$ amount of time (otherwise it would get rejected). So, the total waiting time
for this job is at most $\frac{4 \alpha^2 T_u}{\eps^2 w_j} \leq \frac{8 \alpha^2 \opt}{\eps^2 w_j}.$ Thus, the scheduling algorithm is $O(1/\eps^4)$-competitive.

Further in a phase $u$, the total weight of jobs rejected by $\B(T_u)$ is at most $O(\eps)$-times the total weight processed by $\B(T_u)$ -- this follows from
the analysis for algorithm $\B$. The fact that the queue sizes in $\A(T_u)$ are twice the estimate from Theorem~\ref{thmf:A} (because of the effect of
previous phases) only doubles the weight of rejected jobs. This proves the theorem.
\end{proof}

\section{Extension to {\wtdlinfgen}}
\label{secf:extension}

We now extend our result to the {\wtdlinfgen} problem. Recall that in this problem a 
a job $j$ has two weights associated with it, the rejection-weight $\rw_j$ and flow-time-weight $\wf_j$; the first one is used for counting the rejection weight
of rejected jobs, while the second one is used in the weighted flow-time expression.

It turns out that almost all the details for the algorithm $\A$ carry over with cosmetic changes in notation to this problem as well, however the algorithm $\B$
needs some change; in particular Lemma~\ref{lemf:waste} cannot be applied as it because this is the only place where we critically need the fact that the 
two weights are same. We now outline the modified algorithm and then the changes that are needed in the analysis. As before we shall assume that 
the offline optimum value $\opt$ is known -- the details for removing this assumption are exactly as in the case of {\wtdlinf} problem. 

\vspace*{1 mm}
\noindent
{\bf Algorithm $\A$}
We define the notion of weight class and density class for each of these two weights. Again, assume wlog that both the weights are powers of 2. 
We say that a job $j$ is of rejection-weight class $\rw$ if $\rw_j 
= 2^{\rw}$. Define flow-time-weight class similarly. The rejection-density of a job is defined as $\rw_j/p_j$; and we say  that  its
rejection-density class is $\dr$ if its rejection-density lies in $[2^{\dr},2^{\dr+1})$. Define flow-time-density and flow-time-density class similarly. We now
go over the definitions that were used in defining $\A$ and mention the changes in them. 

A job $j$ is said to be of type $(\wf,\dr)$ if its rejection-density class is $\dr$ and flow-time-weight class is $\wf$. 
For a machine $i$, time $t$, and pair $(\wf,\dr)$, let
$Q_{i,\wf,\dr}(t)$ denote the jobs of type $(\wf,\dr)$ waiting in the queue of machine $i$ at time $t$; and define the
$\load_{i,\wf, \dr}(t)$ as the total weighted remaining processing time of the jobs in $Q_{i,\wf,\dr}(t)$ -- if a job $j$ has remaining
processing time $p_j'$, then its {\em  weighted remaining processing time} is defined as $\wf_j p_j'$. When a job $j$ of type 
$(\wf, \dr)$ arrives at time $t$, we dispatch it to the machine $i \in S_j$ for which $\load_{i,\wf,\dr}(t)$ is minimum, unless for all $i \in S_j,$ 
 $\load_{i,\wf,\dr}(t) + p_j.2^{\wf} \geq \alpha^2 \cdot \opt$. If the latter case happens,  we reject this job.

It remains to specify which job is processed at any time by a machine.
For a time $t$ and machine $i$, let $(w^f_i(t), d^r_i(t))$ be the pair $(\wf,\dr)$ with the highest $2^{\dr} \cdot \load_{i,\wf,\dr}(t)$ value.
  We process the earliest released job from the queue $Q_{i,w^f_{i}(t), d^r_{i}(t)}(t)$ on machine $i$ at time $t$. Assume a fixed rule of breaking ties.  

\vspace*{1mm}
\noindent
{\bf Algorithm {$\B$}: }\ Now we describe the modified algorithm $\B$. 
When a job $j$ arrives at time $t$, it is dispatched according to $\A$: if $\A$ rejects this job, $\B$ also rejects
it; and if $\A$ dispatched it to machine $i$, then $\B$ also dispatches this job to $i$. Now, we describe the processing policy for a fixed machine $i$. Consider a time $t$. Let $d^\A(t)$
denote the rejection-density class of the job processed by $\A$ at time $t$. Then, $\B$ processes the following job at time $t$: if there is a job of rejection-density class 
higher than $d^\A(t) + 2 \logeps$ in the queue of
machine $i$ at time $t$, then $\B$ processes any such job; otherwise it processes the job of rejection-density class $d^\A(t)$ with the highest flow-time-weight class (it prefers the earliest released job in case of ties).
 Also note that if the second
case happens and there is no job of rejection-density class $d^\A(t)$ in the queue of machine $i$ at time $t$ (in $\B$), we can process any job at this time.

The algorithm $\B$ may reject some more jobs. There are two kinds of rejections: (i) immediate rejection: for each triplet $(\wf,\dr,p)$, $\B$ rejects every $\left(
\frac{1}{\eps} \right)^{th}$ job of type $(\wf,\dr)$ and size-class $p$ dispatched to it -- note that these jobs are rejected as soon as 
they are released (the algorithm $\A$ is immediate dispatch), (ii) delayed rejection: 
for every flow-time-weight class $\wf$, we divide the time line into segments of length $\frac{6 (\alpha^2+2) \opt}{\eps^4 2^{\wf}}$.
Suppose a job of flow-time-weight class $\wf$ gets released during such a segment $S$, and let $S'$ be the segment immediately to the right of $S$. If
the job does not complete processing by the end of $S'$, the algorithm $\B$ rejects the job. This completes the description of $\B$.

Now we outline how the analyses of these two algorithms change.

\noindent
{\bf Analysis of $\A$:} 
The goal is to prove the following extension of Theorem~\ref{thmf:A}.
\begin{theorem}
\label{thmf:Agen}
The algorithm $\A$ rejects jobs of total rejection-weight at most $\eps$ times the total rejection-weight of all jobs, and ensures that for any machine $i$, time $t$, and 
pair $(\wf,\dr)$, the total weighted remaining processing time of jobs of type $(\wf,\dr)$ at time $t$ on machine $i$ is at most $\alpha^2 \opt$. Further, $\A$ is an immediate dispatch algorithm which rejects jobs on arrival only.
\end{theorem}
\begin{proof}
We give a brief description of the changes that are needed from the proof of Theorem~\ref{thmf:A}. A weighted machine interval is again a triplet $(I,i,\wf)$,
where $\wf$ corresponds to a flow-time-weight class. Lemma~\ref{lemf:genlower} holds if we replace weight by flow-time-weight (the lemma gives a lower
bound on $\opt$, which does not depend on rejection at all).  Again, we fix a time $t^\star$, and a rejection-density class $\dsr$. The quantities $\alpha$
and $\gamma_l$ are defined as before. For a machine $i$, time $t$, and rejection-density class $\dr$, $\load_{i,\dr}(t)$ is defined as the maximum over
all flow-time-weight class $\wf$, of $\load_{i,\wf,\dr}(t)$. The quantities $\level(j), \jobg{l}{\dsr}$ and $\maxload_{i,\dsr}(t)$ are defined analogously.

The weighted machine intervals $\intg{i}{l}{\dsr}$ are constructed as in Figure~\ref{figf:intwg} (in Step 2(iii), $w$ refers to flow-time-weight class).
Lemma~\ref{lemf:propwg} follows without any change (in the proof, we use $d$ to refer to rejection-density class and $w$ to flow-time-weight class). 
The dual variables are defined as before, and so, Claim~\ref{clmf:wqprocess} follows without any change.

For a parameter $l$ and flow-time-weight class $\wf$, let $\bint{l}{\wf}{\ds}$ is defined as before, 
and the quantities $m_{l,\wf,\dsr}, \len{l,\wf,\dsr}$ are defined analogously. Inequality~(\ref{eqf:lengthBound}) holds without any change
(if we replace $w$ by $\wf$). The same applies to Claim~\ref{clmf:lengthwg} and inequality~(\ref{eqf:lowerB2}). The definitions
$V_{w,d}$ and $W_{w,d}$  get replaced by (note the subtle change below -- the weights in the summation are the flow-time weights in the
definition of $V$ but the rejection-weights in the definition of $W$) : 

$$V_{\wf, \dr}=\sum_{j: \type(j)=(\wf,\dr)} w^f_j p_j, \ \ W_{\wf,\dr} = \sum_{j: \type(j)=(\wf,\dr)} w^r_j. $$

$V_{\wf,\dr}(a,b)$ is defined analogously. Claim~\ref{clf:countwg} goes through without any changes (if we replace $w$ by $\wf$ and $d$ by $\dr$). 
Note that in the proof of this claim, the last line will now read as
$$\frac{2^{\dr}}{2^{\wf}} V_{\wf,\dr} = \sum_{j: \type(j)=(\wf,\dr)} \frac{2^{\dr}}{w^f_j} w^f_j p_j =  
\sum_{j: \type(j)=(\wf,\dr)} 2^{\dr} p_j \leq  \sum_{j: \type(j)=(\wf,\dr)} w^r_j  = W_{\wf,\dr}.$$

Proof of Lemma~\ref{lemf:finalA} now follows without any changes. 
\end{proof}

\noindent
{\bf Analysis of $\B$:}   We give some intuition about where the analysis differs from that in the case of \wtdlinf. As noted earlier, in the case of \wtdlinf,
we needed to bound the volume of bad time slots, i.e., time where $\B$ processes a job, but then decides to reject it later. For this, we crucially needed
the fact that the two kind of weights are same. However, we do not have this luxury anymore. This is where we use the immediate rejection -- if volume
$V$ of jobs (of a certain rejection density class $d$) are dispatched to a machine $i$, about $\eps V$ of this will get immediately rejected. Now, it may
happen that $\B$ rejects the remaining jobs after processing them to almost completion, but even then the volume of bad jobs will remain at most
$(1-\eps)V$. In the remaining $\eps V$ time slots (meant for such jobs), $\B$ must process very high density jobs, and so will be able to bound the
weight of rejected jobs.

We now proceed with the analysis of $\B$. 
We only need to bound the total rejection weight of jobs which are rejected.
The case of immediate rejection is easy.

\begin{claim}
\label{clf:easyrej}
The total rejection weight of jobs which are rejected by $\B$ using immediate reject is at most $4 \eps$ times the total rejection weight of all jobs. 
\end{claim}
\begin{proof}
Consider a job of flow-time weight class $\wf$, rejection density class $\dr$ and size class $p$. Its rejection weight  is at least $2^p \cdot 2^{\dr}$
and at most $2^{p+1} \cdot 2^{\dr+1}$. In other words, any two such jobs will differ in their rejection weight by a factor 4 only. Since $\B$ rejects
such jobs after every $1/\eps$ arrivals, the claim follows. 
\end{proof}

The case of delayed rejects is more non-trivial. Again, we fix a machine $\is$ and rejection density class $\dsr$. For ease of notation, we shall remove the
subscripts involving $\is$ and $\dsr$. Let $\jrej$ denote the set of jobs of rejection density class $\dsr$ which are rejected (using delayed reject) by
$\B$. We  shall again define a set of disjoint  intervals  but their definition is more tricky. For a flow-time weight class $\wf$, we 
can divide the time line into disjoint set of intervals $\I(\wf)$ such that any interval $I \in \I(\wf)$ satisfies: (i) at any point of time $t \in I$, there
is a job of type $(w, \dsr), w \geq \wf$, in the queue of machine $\is$ in $\B$, and (ii) at least one job of type $(\wf, \ds)$ in $\jrej$ is released during $I$. 
In other words, we first divide the time line into disjoint maximal intervals where we have a job of type $(w, \dsr), w \geq \wf,$ in the queue of machine $\is$.
From these intervals, we pick out those where a job in $\jrej$ of type $(\wf, \dsr)$ is released. 

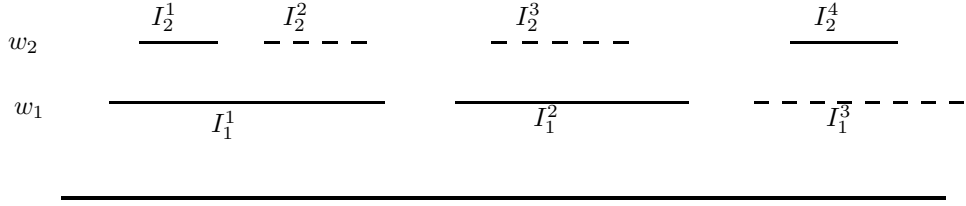
\begin{figure}
\begin{center}
\setlength{\unitlength}{1973sp}%
\begingroup\makeatletter\ifx\SetFigFont\undefined%
\gdef\SetFigFont#1#2#3#4#5{%
  \reset@font\fontsize{#1}{#2pt}%
  \fontfamily{#3}\fontseries{#4}\fontshape{#5}%
  \selectfont}%
\fi\endgroup%
\begin{picture}(12037,2493)(511,-3394)
\put(10801,-2461){\makebox(0,0)[lb]{\smash{{\SetFigFont{10}{12.0}{\rmdefault}{\mddefault}{\updefault}{\color[rgb]{0,0,0}$I_1^3$}%
}}}}
\thicklines
{\color[rgb]{0,0,0}\put(1801,-2161){\line( 1, 0){3450}}
}%
{\color[rgb]{0,0,0}\put(6151,-2161){\line( 1, 0){2925}}
}%
{\color[rgb]{0,0,0}\put(2176,-1411){\line( 1, 0){975}}
}%
{\color[rgb]{0,0,0}\put(10351,-1411){\line( 1, 0){1350}}
}%
{\color[rgb]{0,0,0}\multiput(9901,-2161)(350.00000,0.00000){8}{\line( 1, 0){175.000}}
}%
{\color[rgb]{0,0,0}\multiput(3751,-1411)(364.28571,0.00000){4}{\line( 1, 0){182.143}}
}%
{\color[rgb]{0,0,0}\multiput(6601,-1411)(383.33333,0.00000){5}{\line( 1, 0){191.667}}
}%
\put(601,-2311){\makebox(0,0)[lb]{\smash{{\SetFigFont{10}{12.0}{\rmdefault}{\mddefault}{\updefault}{\color[rgb]{0,0,0}$w_1$}%
}}}}
\put(526,-1486){\makebox(0,0)[lb]{\smash{{\SetFigFont{10}{12.0}{\rmdefault}{\mddefault}{\updefault}{\color[rgb]{0,0,0}$w_2$}%
}}}}
\put(3976,-1186){\makebox(0,0)[lb]{\smash{{\SetFigFont{10}{12.0}{\rmdefault}{\mddefault}{\updefault}{\color[rgb]{0,0,0}$I_2^2$}}}}}
\put(10651,-1186){\makebox(0,0)[lb]{\smash{{\SetFigFont{10}{12.0}{\rmdefault}{\mddefault}{\updefault}{\color[rgb]{0,0,0}$I_2^4$}%
}}}}
\put(6901,-1186){\makebox(0,0)[lb]{\smash{{\SetFigFont{10}{12.0}{\rmdefault}{\mddefault}{\updefault}{\color[rgb]{0,0,0}$I_2^3$}%
}}}}
\put(2326,-1186){\makebox(0,0)[lb]{\smash{{\SetFigFont{10}{12.0}{\rmdefault}{\mddefault}{\updefault}{\color[rgb]{0,0,0}$I_2^1$}%
}}}}
\put(3076,-2536){\makebox(0,0)[lb]{\smash{{\SetFigFont{10}{12.0}{\rmdefault}{\mddefault}{\updefault}{\color[rgb]{0,0,0}$I_1^1$}%
}}}}
\put(7126,-2461){\makebox(0,0)[lb]{\smash{{\SetFigFont{10}{12.0}{\rmdefault}{\mddefault}{\updefault}{\color[rgb]{0,0,0}$I_1^2$}%
}}}}
{\color[rgb]{0,0,0}\put(1201,-3361){\line( 1, 0){11100}}
}%
\end{picture}%
\end{center}
\caption{The intervals $I_1^1, I_1^2, I_1^3$ are maximal intervals where a job of flow-time weight class at least $w_1$ is waiting in the queue. Out of these
$\I(w_1)=\{I_1^1, I_1^2\}$ because these are the intervals where a delayed rejected job of type $(w_1, \dsr)$ is released. Similarly, for a 
flow-time weight class $w_2, w_2 \geq w_1$, $\I(w_2)=\{I_2^1, I_2^4\}$. Finally, $\I=\{I_1^1, I_1^2, I_2^4.\}$}
\label{figf:example2}
\end{figure}

Note that if $I \in \I(\wf_1), I' \in \I(\wf_2)$, then either $I$ and $I'$ are disjoint or one is contained in the other. Let $\I$ denote the (containment wise)
maximal intervals in $\cup_{\wf} \I(\wf)$, i.e., any two intervals in $\I$ are disjoint and no interval in $\I$ is contained inside any other interval in 
$\cup_{\wf} \I(\wf)$ (see Figure~\ref{figf:example2}). 
Fix an interval $I \in \I$ for rest of the discussion. 
Let $\jrej(I)$ be the set of jobs in $\jrej$ which are released during $I$.

Let $w_{\min}(I)$ be the smallest flow-time weight class of a job in $\jrej(I)$ -- observe that $I \in \I(\wmin(I))$. 
Further, at every point of time in $I$, there is a job of type $(w, \dsr), w \geq \wmin(I)$ in the queue of machine $\is$
in the algorithm $\B$.
 The maximality of $I$ shows that 
there is no such job in the queue of machine $\is$ just before the left end-point of $I$.

For a size class $p$ and flow-time weight class $\wf$, let $P^{\A}_{\dsr, p,\wf}(I)$
denote the total volume of time in $I$ during which $\A$ processes jobs of type $(\wf, \dsr)$ and size class $p$
  (on the machine $\is$). 
Let $J^{\A}_{\dsr,p,\wf}(I)$ denote the set of jobs of type $(\wf, \dsr)$ and size-class $p$
 which are dispatched by $\A$ to $\is$ during $I$,  and $n_{\dsr, p, \wf}$ denotes the cardinality of
this set. Note that $\A$ will either process these jobs during $I$ or they will end up in the queue of $\is$ at the end of $I$. 

First observe that for any $\wf$, 
\begin{eqnarray}
\label{eqnf:proc}
\sum_p \sum_{j \in J^{\A}_{\dsr,p,\wf}(I)} p_j \leq  \sum_p P^{\A}_{\dsr, p,\wf}(I) + \alpha^2 \opt/2^{\wf},
\end{eqnarray}
because any job in the LHS will either be processed by $\A$ during $I$ or will end up in the queue of type $(\wf,\dr)$ jobs at the end of $\is$, and the latter
part can have total remaining processing volume at most $\alpha^2 \opt/2^{\wf}$ (Theorem~\ref{thmf:Agen}). 

%
%

\begin{claim}
\label{clf:Bproc}
The total volume of time in $I$ during which $\B$ processes jobs of type $(\wf,\dsr), \wf \geq \wmin(I),$  is at most
$$ \left( 1 - \eps/2 \right)  P^{\A}_{\dsr}(I) +  2(\alpha^2 +2) \opt/2^{\wmin(I)} , $$
where $P^{\A}_{\dsr}(I)$ denotes $\sum_p \sum_{\wf \geq \wmin(I)} P^{\A}_{\dsr, p,\wf}(I) $, i.e., the total volume of time in $I$ during
which $\A$ processes jobs of rejection-density class $\dsr$ and weight-flow-time class $\wmin(I)$ or higher. 
\end{claim}

\begin{proof}
Observe that any job of type $(\wf, \dsr)$, $\wf \geq \wmin(I)$, processed  by $\B$ during $I$ (on machine $\is$) has to either:
(i) appear in the queue of $\B$ on machine $\is$ at the beginning of $I$, or (ii) dispatched to $\is$ by $\A$ during $I$ and not get immediately rejected
by $\B$. By definition of $I$ (maximality property), there is no job in (i). We now estimate (ii). Note that for any  flow-time weight class $\wf$ and size class $p$, 
$\B$ will immediately reject at least 
$ \sum_p (\eps \cdot n_{\dsr, p, \wf} - 1)$ jobs from $J^{\A}_{\dsr,p,\wf}(I)$, and so, the total volume of such jobs rejected by $\B$ is at least 
$$ \sum_p (\eps \cdot n_{\dsr, p, \wf} - 1) \cdot 2^p \geq  \sum_p \eps \cdot n_{\dsr, p, \wf} \cdot 2^p - 2\opt/2^{\wf} 
\geq \eps/2 \cdot \sum_p \sum_{j \in J^{\A}_{\dsr,p,\wf}(I)} p_j  - 2\opt/2^{\wf}, $$
where the first inequality follows from the fact that the maximum processing time of any job of flow-time-weight class $\wf$ is at most $\opt/2^{\wf}$. 
Inequality~(\ref{eqnf:proc}) and the above inequality now imply that  the total volume of time during $I$ at which $\B$ processes a job of type $(\wf, \dsr)$ is at most
\begin{eqnarray*}
\sum_p \sum_{j \in J^{\A}_{\dsr,p,\wf}(I)} p_j -   \eps/2 \cdot \sum_p \sum_{j \in J^{\A}_{\dsr,p,\wf}(I)} p_j  + 2\opt/2^{\wf}
\leq \left( 1 - \eps/2 \right) \sum_p P^{\A}_{\dsr, p,\wf}(I) + (\alpha^2 +2) \opt/2^{\wf} 
\end{eqnarray*}
Summing over all $\wf \geq \wmin(I)$, we get the desired result.
\end{proof}

It follows from the above claim that there must be at least 
\begin{eqnarray}
\label{eqf:countgen}
P^{\A}_{\dsr}(I) - \left( \left( 1 - \eps/2 \right)  P^{\A}_{\dsr}(I) +  2(\alpha^2 +2) \opt/2^{\wmin(I)} \right)
= \eps/2 \cdot P^{\A}_{\dsr}(I) - 2(\alpha^2 +2) \opt/2^{\wmin(I)} 
\end{eqnarray}
 volume of time during $I$ at which $\B$ is performing jobs of rejection-density class $\ds+2 \logeps$ or higher -- indeed,
there is always a job of type $(\wf, \dsr), \wf \geq \wmin(I)$ in the queue of $\is$ during $I$, and yet, $\B$ processes such jobs during
 $ \left( 1 - \eps/2 \right)  P^{\A}_{\dsr}(I) +  2(\alpha^2 +2) \opt/2^{\wmin(I)}  $ out of the possible $P^{\A}_{\dsr}(I) $ 
volume (during which $\A$ processes such jobs in $I$). 
Now, define an indicator variable $\bone{\B}{\is}{\geq \ds}(t)$ which is 1 if $\B$ processes job of rejection-density class at least $\ds$ at time $t$ on $\is$.
\begin{lemma}
\label{lemf:countgen}
The total processing time of $\jrej(I)$, the jobs of rejection-density class $\dsr$ which get delayed rejected by $\B$ during $I$, is at most
\begin{eqnarray}
\label{eqf:tot}
\frac{2}{\eps} \int_{t \in I} \bone{\B}{\is}{\geq \ds+ 2\logeps} dt + \eps^3 \int_{t \in I} \bone{\B}{\is}{\geq \ds- 2\logeps},
\end{eqnarray}
\end{lemma}
\begin{proof}
Inequality~(\ref{eqf:countgen}) implies that the first term in the summation above is at least
$$2/\eps \left(\eps/2 \cdot P^{\A}_{\dsr}(I) - 2(\alpha^2 +2) \opt/2^{\wmin(I)} \right) = P^{\A}_{\dsr}(I) - \frac{4(\alpha^2 +2) \opt}{\eps \cdot 2^{\wmin(I)}},$$
and the second term is at least (recall that at all time during $I$, $\B$ will be processing a job of rejection density class at least $\dsr-2\logeps$)
$$\eps^3 \cdot \length(I) \geq \frac{6(\alpha^2 +2) \opt}{\eps \cdot 2^{\wmin(I)}},$$
because $I$ contains a job of rejection-weight $\wmin(I)$, and so must be as long as one segment corresponding to this rejection-weight class (according to
the description of algorithm $\B$). Therefore, the expression in~(\ref{eqf:tot}) is
at least $$P^{\A}_{\dsr}(I)+\frac{2 \alpha^2 \opt}{\eps \cdot 2^{\wmin(I)}}, $$ which is at least the total processing size of $\jrej(I)$ (using 
inequality~(\ref{eqnf:proc}) and summing over all $\wf \geq \wmin(I)$). 
\end{proof}

Rest of the argument follows as in the case of Theorem~\ref{thmf:B}
-- we sum the expression in~(\ref{eqf:tot}) over all $I \in \I$, and then over all rejection-density classes $\dsr$ to show that the total rejection weight of rejected
jobs is within $\eps$ fraction of the total rejection weight of all jobs. 
 Thus, we have shown
\begin{theorem}
\label{thmf:Bgen}
The algorithm $\B$ is $O(1/\eps^6)$-competitive algorithm and rejects jobs of total rejection weight at most $O(\eps)$-times the total rejection-weight 
of all jobs. 
\end{theorem}

\section{Some Lower Bounds}
\label{secf:lower}
In this section, we show results on some lower bounds of the {\loadbal} and the {\linf} problems.
The first result shows that for {\loadbal}, the trade-off that we obtain between competitive ratio and the fraction
of jobs rejected is nearly optimal.
\begin{lemma}
\label{lemf:lower-load}
Given a parameter $\eps$, and an (deterministic)
online immediate dispatch  algorithm $\A$ for the {\loadbal} problem,
there is an input $\I(\eps)$ consisting of unit size jobs
such that $\A$ rejects at most $\eps$-fraction
of the jobs and the competitive ratio of $\A$  on $\I(\eps)$
is at least $\Omega \left( \log \left( \frac{1}{\eps} \right) \right)$.
\end{lemma}
\begin{proof}
The input $\I(\eps)$ has $m$ machines, and will have at most $2m$ jobs. So,
 the algorithm
can reject at most $2 \eps m$ jobs. The jobs are released in several phases.
At the beginning of phase $l$,
we have a set $M_l$ of machines on which load is at least
$l$. We shall use $m_l$ to denote $|M_l|$.

This is clearly true for $l=0$ with $M_0$ being the initial set of machines and $m_0 = m$.
Suppose the invariant is true at the beginning of a phase $l$. During phase $l$, we shall
assume wlog that the algorithm first dispatches {\em all} the jobs released during this phase, and
then rejects some of them. This will not change the competitive ratio of the algorithm because we
will only look at the end of a particular phase.
We partition the machines in $M_l$ into $\frac{m_l}{2}$
disjoint pairs. For each such pair $i,i'$, we release $2$ jobs which can only go on these two machines.
Without loss of generality, assume that the algorithm $\A$ dispatches at least one of these two jobs to $i$.
Then we add $i$ to a set $X$. Thus, $X$ has $\frac{m_l}{2}$ machines. Assuming $m_l \geq 8 \eps m$, we can
reject at most $2 \eps m \leq \frac{|X|}{2}$ jobs dispatched to the machines in $X$ during this phase.
Thus, at least half of the machines in $X$ get at least one job during this phase
 -- we let $M_{l+1}$ be these machines.
Observe that $\frac{m_l}{4} \leq m_{l+1} \leq \frac{m_l}{2}$. Further the load on a machine in $M_{l+1}$ is at least $l+1$.
Thus, the invariant holds at the end of this phase as well.

Note that we require $m_l \geq 8 \eps m$. This will hold if we have only $\left( \log
\left( \frac{1}{\eps} \right) \right)$ phases. It is easy to check that the
optimal load is at most $2$ -- during phase $l$, when we send $2$ jobs to a pair of machines $i,i'$,
the offline  schedule sends these jobs to the machine which does not get added to $X$.
Further, the total number of jobs released is at most $\sum_l m_l \leq 2m$.
This completes the proof of the lemma.
\end{proof}

The next result shows that for arbitrary processing times, an immediate dispatch and immediate reject algorithm
for the {\loadbal} problem will incur high competitive ratio.
\begin{theorem}
\label{thmf:lowerunbounded}
Any online algorithm $\A$ for the {\loadbal} problem which satisfies immediate dispatch and immediate reject has unbounded
competitive ratio, even if it can reject $\eps$-fraction of the jobs.
\end{theorem}
\begin{proof}
The proof is very similar to that of Lemma~\ref{lemf:lower-load}.
However, in subsequent phases, jobs sizes will
start decreasing, and so, the number of phases will not be bounded by a function of $\eps$. We shall maintain
the following invariants for all phases $l=0, 1, 2, \ldots:$
\begin{itemize}
\item At the beginning of phase $l$, there will be  a set $M_l$ of $m_l = \frac{m}{4^l}$ machines such that the
load on each of them (in the schedule produced by algorithm $\A$)
 will be at least $\frac{l}{2}$. Here $m$ is the initial number of machines.
\item During phase $l-1$, we shall release $2^l \cdot m$ jobs each of length $\frac{1}{8^l}$.
\end{itemize}
These invariants hold at $l=0$: $M_0$ is the initial set of machines. Suppose these properties hold for some
 $l$. Again, we assume wlog that the algorithm first dispatches all the jobs released in a phase, and then
rejects some of them. During phase $l$, we pair up the machines in $M_l$ into $\frac{m_l}{2}$ pairs. For each such pair of
machines $i,i'$, we release $2 \cdot 8^l$ jobs, each of length $\frac{1}{8^l}$, which can only be processed
by these two machines. Assume without loss of generality  that the algorithm $\A$ dispatches at least $8^l$
of these jobs to $i$ -- we add $i$ to a set $X$. Hence, $X$ will be a set of $\frac{m_l}{2}$ machines.

Note that the total number of jobs released in phase $l$ is $m_l \cdot 8^l = m \cdot 2^l$. Therefore, the
total number jobs released so far is at most $m \cdot 2^{l+1}$, and so, $\A$ can reject at most $\eps
\cdot m \cdot 2^{l+1} = 2 \eps \cdot m_l 8^l $ jobs released during this phase -- note that $\A$ is
not allowed to reject jobs released in previous phases during phase $l$. This implies that at least half
of the machines in $X$ will receive at least $\frac{8^l}{2}$ jobs during this phase -- otherwise, the
total number of rejected jobs will be at least $\frac{|X| \cdot 8^l}{4} = \frac{1}{8} \cdot m_l \cdot 8^l$,
which will be a contradiction (assuming $\eps < 1/16$). Since $|X|/2 \geq m_l/4$, we pick $M_{l+1}$ to
be $m_l/4$ such machines. Note that the total load on these machines is at least $\frac{l}{2} + \frac{1}{8^l}
\cdot \frac{8^l}{2} = \frac{l+1}{2}$. This proves the invariant for $l+1$.

It is also easy to see that the optimal offline load on any machine is at most 2 (using the same argument
as in the proof of Lemma~\ref{lemf:lower-load}. Since the number of phases can be arbitrarily large (depending
on the value of $m$), we see that the competitive ratio of $\A$ is unbounded.
\end{proof}

Next we give a strengthening of Lemma~\ref{lemf:lower-load} for the {\linf} problem.
\begin{lemma}
\label{lemf:lower-linf}
Given a parameter $\eps$, and an (deterministic)
online immediate dispatch  algorithm $\A$ for the {\linf} problem,
there is an input consisting of unit size jobs
such that $\A$ rejects at most $\eps$-fraction
of the jobs and the competitive ratio of $\A$  is $\Omega \left( \frac{1}{\eps} \right).$
\end{lemma}

\begin{proof}
In our input, we shall release jobs at the end of a time interval $I$. We assume that the algorithm $\A$ dispatches all the jobs
released during $I$, and rejects jobs only at the end of $I$. This will be without loss of generality: we will only consider
the queue size on a machine at the end of the interval $I$. If $\A$ was rejecting any job during $I$, then it could instead
dispatch this job arbitrarily and reject it at the end of $I$. This will not affect the queue length of machines at the end
of $I$.

We shall use the following gadget while building the input.
\begin{claim}
\label{clf:gadget}
Suppose at some time $t$, the algorithm $\A$ is given as input, a set $M$ of $m+1$ machines with $i$ unit sized jobs in
the queue of the $i^{th}$ machine, $i=0, \ldots, m$. Then one can release $m+3$
unit size jobs during the interval $[t, t+2]$
such that at time $t+2$, the load (queue size) on these machines (in some other order) are $0,1,2, \ldots, m-1, m+1$.

Further, there is an offline algorithm which given the machines in $M$ with unit load at time $t$ and
the same input as above during $[t, t+2]$ ends up with unit load on all machines at time $t+2$. The maximum
load on any machine during this interval is $3$.
\end{claim}
 \begin{proof}
We will release several jobs at the same time in a {\em sequence}, i.e., the next job will be released only after the
current one is dispatched by $\A$. 
We label the machines $M_0, \ldots, M_m$, with $M_i$ having load $i$ at time $t$ in $\A$. All jobs have unit size.
The procedure for releasing jobs is described in Figure~\ref{figf:dispatch}.
We first observe that at the beginning of Step~3(i), both $M_{i-1}$ and $M_i$ have load $i$. Hence, at the end of time $t$,
machine $M_i$ has load $i+1$. Due to Step 4, at the end of time $t+1$, $M_m$ continues to have load $m+1$, but
now, $M_i$ has load $i$ for all $i=0, \ldots, m-1$. Further, no change in load happens due to Step 5, because
$M_0$ already had 0 load. This step is needed to argue about the offline algorithm.

Finally, it is easy to check that there is an offline schedule with the desired properties. In Step~3(ii),
the job $j_i$ is dispatched to $M_{i-1}$. Thus, at the end of time $t$, $M_0$ has 3 jobs in its queue, $M_1, \ldots, M_{m-1}$
get 2 jobs, and $M_m$ has one. Step 4 ensures that at the end of time $t+1$, $M_0$ has 2 jobs in its queue, and the remaining
machines have one jobs each in their queue. Finally, in Step 5, $M_0$ at the end of time $t+2$ also ends up with one job in its queue.
This proves the desired result.

\begin{figure}[ht]
   \begin{center}
     \begin{boxedminipage}{6.5in}
          {\bf IncreaseLoad:} \medskip\\
        \sp \sp 1.   At time $t$, release a job $j_0$ which can only be processed by $M_0$. \\
         \sp \sp 3. {\bf For} $i=1$ to $m$ {\bf do} \\
         \sp \sp \sp \sp (i) Release a job $j_i$ at time $t$ which can be processed by $\{M_{i-1}, M_i\}$ \\
         \sp \sp \sp \sp (ii) Assume (upto relabeling) that $\A$ dispatches $j_i$ to $M_{i}$. \\
         \sp \sp 4. At time $t+1$, release one job $j_m'$ which can only go on machine $M_m$. \\
         \sp \sp 5. At time $t+2$, release one job $j_i''$ for $i=1, \ldots, m$, such that \\
         \sp \sp \sp \sp $j_i''$ can be processed by $M_i$ only. \\
      \end{boxedminipage}
       \caption{Dispatching jobs for Claim~\ref{clf:gadget}}
       \label{figf:dispatch}
       \end{center}
 \end{figure}

\end{proof}

In our input, we shall have $\Delta$ machines, where the parameter $\Delta$ will be specified later. The jobs are
released in $\Delta-2$ stages. In the beginning of stage $l$, following invariants are satisfied: (i) In the schedule
produced by $\A$, there are $l$ machines with load $1, 2, \ldots, l$, and rest of the machines have 0 load, (ii)
In the offline schedule, load is 1 on all machines. Further, the maximum load on any machine till the beginning
of stage $l$ was 3.

 Clearly,  this is true at the beginning of stage 0. Suppose
this holds true for $l$ -- let $t_l$ denote the time at the beginning of stage $l$. We iteratively call the procedure
{\bf IncreaseLoad} defined in Claim~\ref{clf:gadget}. The procedure is described in Figure~\ref{figf:lower}.
Assume that machine $M_i$ has load $i$ for $i=1, \ldots, l$.  The load on other machines remains 0.
 The procedure uses two other  machines, which we call $M_0$
and $M_0'$ (note that we have at least $l+2$ machines).  During the procedure, we may keep a machine
{\em busy} during a time period. This essentially means that we release one (unit size) at every unit time step during
this time period -- hence, the load on this machine does not change at all.

\begin{figure}[ht]
   \begin{center}
     \begin{boxedminipage}{6.5in}
          {\bf ReleaseJobs(Stage l):} \medskip\\
        \sp \sp 1. Initialize time $t \leftarrow t_l$ \\
        \sp \sp 2. {\bf For} $i= l$ downto $1$ {\bf do} \\
         \sp \sp \sp \sp (i) Machines $M_0, \ldots, M_i$ have load $0, \ldots, i$ (upto reordering) and \\
         \sp \sp \sp \sp \sp \sp machines $M_{i+1}, \ldots, M_l$ have load $i+2, \ldots, l+1$ (upto reordering) \\
         \sp \sp \sp \sp (ii) Call {\bf IncreaseLoad} at time $t$ with machines $M_0, \ldots, M_i$ -- these machines \\
         \sp \sp \sp \sp \sp \sp \sp \sp \sp  get load $0,1,2,\ldots,i-1, i+1$ (upto reordering) at time $t+2$ \\
         \sp \sp \sp \sp (iii) {\bf For} $j=i+1$ to $l$ \\
         \sp \sp \sp \sp \sp \sp \sp \sp \sp Keep machine $M_j$ busy during $[t,t+2]$. \\
         \sp \sp \sp \sp (iv) $t \leftarrow t+2$. \\
         \sp \sp 3. Upto reordering $M_1, \ldots, M_{l}$ have load $2, \ldots, l+1$, and $M_0$ has load 0. \\
         \sp \sp \sp \sp (i) At time $t$ release a job which can be processed by $M_0$ and $M_0'$ only, \\
         \sp \sp \sp \sp \sp \sp \sp \sp assume $\A$ dispatches this job to $M_0$ (upto reordering). \\
         \sp \sp \sp \sp (ii) Keep $M_0, \ldots, M_l$ busy during $[t,t+1]$.
      \end{boxedminipage}
       \caption{Jobs released in a stage}
       \label{figf:lower}
       \end{center}
 \end{figure}

It is easy to check that at the end of this procedure, we have $l+1$ machines with load $1,2 \ldots, l+1$
respectively. Suppose there is an offline algorithm in which all machines have load 1 at time $t_i$. Then
Claim~\ref{clf:gadget} implies that during Step 2, the maximum load on any machine stays 3, and we end up with
1 load on all machines. In Step~3(i), the job released is sent to $M_0'$ by the offline schedule, and
then during Step~3(ii), $M_0'$ processes this job to end up with 1 load. This proves that the invariants
hold for the next stage as well. Note that the total number of jobs released during this stage (using
Claim~\ref{clf:gadget}) are at most:
$$ 1 + \sum_{i=1}^l \left[ (i+3) + 2(l-i) \right] + 1 + (l+1) \leq \frac{3}{2} l^2. $$
Thus, the total number of jobs released till $\Delta-2$ stages are at most $\frac{\Delta^3}{4}$. At the end
of final stage, we have at $\Delta-1$ machines with load $0,1, \ldots, \Delta-1$. Now, the online
algorithm can delete at most $\eps \cdot \frac{\Delta^3}{4} \leq \frac{\Delta^2}{8}$ jobs (assuming
$\Delta = \frac{1}{2 \eps}$). Thus, we will still have at least one machine with total load at least $\frac{\Delta}{2}
= \frac{1}{4 \eps}$ -- indeed, otherwise the total number of deleted jobs will be at least
$$ 1 + 2 + \cdots + \frac{\Delta-1}{2} \geq \frac{\Delta^2}{8}.$$ This proves the lemma. Observe that here,
the input size is constrained by a function of $\eps$ -- but we can make it independent of $\eps$ by
 taking multiple copies of the above  construction in parallel.
\end{proof}

\section{Conclusion and Open Problems}
In this paper, we proposed a new model for avoiding the pessimistic bounds arising from competitive analysis of online algorithms for scheduling problems. 
We could give constant-competitive algorithms for load balancing and minimizing maximum weighted flow-time problems in this model, even though such 
results cannot be obtained in the speed augmentation model. It is not difficult to show that if there is a single machine, then a policy which rejects 
every $(\frac{1}{\eps})^{th}$ job (for each weight-class and job size class) and follows HDF rule has competitive ratio within a constant of that when 
we allow the machine $(1+\eps)$-speed augmentation. Hence, the results which give immediate dispatch algorithms with 
competitive ratio of $\frac{p}{\eps^{O(1)}}$ for minimizing
$\ell_p$ norm of flow-time in the unrelated machines model with speed augmentation (~\cite{AnandGK12,ImM11}) can be proved here as well (with an extra $1/\eps$ factor loss
in competitive ratio). It is an interesting problem to give  algorithms for minimizing $\ell_p$ norm of flow-time in the rejection model with 
competitive ratio independent of $p$. More generally, we feel that more interesting results can be given for other online scheduling problems in this model. 

\bibliographystyle{plain}
\bibliography{reject_modified2}

\end{document}